%% file: data_sway.tex
\documentclass[sigconf,screen,authorversion]{acmart}
\AtBeginDocument{%
  }

\copyrightyear{2026}
\acmYear{2026}
\setcopyright{cc}
\setcctype{by-nc-nd}
\acmConference[DIS '26]{Designing Interactive Systems Conference}{June 13--17, 2026}{Singapore, Singapore}
\acmBooktitle{Designing Interactive Systems Conference (DIS '26), June 13--17, 2026, Singapore, Singapore}
\acmDOI{10.1145/3800645.3813048}
\acmISBN{979-8-4007-2563-0/2026/06}

\input{meta/commands}

\begin{document}

\title{DataZephyr: Authoring SVG Animation for Metaphoric Visualization with GenAI}
\title{DataSway: A Mixed-Initiative Interface for Animating SVG-based Metaphoric Data Visualizations}
\title{DataSway: A Composite Framework to Animating Metaphoric Data Visualizations}
\title{Towards Controllable SVG Animation Creation for Metaphoric Data Visualization}
\title{DataSway: Animate Metaphoric Visualization by Coordinating Keyframes for Data Elements}

\title{DataSway: Coordinating Animation Clips of Data Elements in Dynamic Metaphoric Visualization}

\title{DataSway: Vivifying Metaphoric Visualization with Animation Clip Generation and Coordination}

\input{meta/authors}

\begin{abstract}
 \input{meta/abs}
\end{abstract}

\input{meta/keywords}

\begin{teaserfigure}
\centering
  \includegraphics[width=\textwidth]{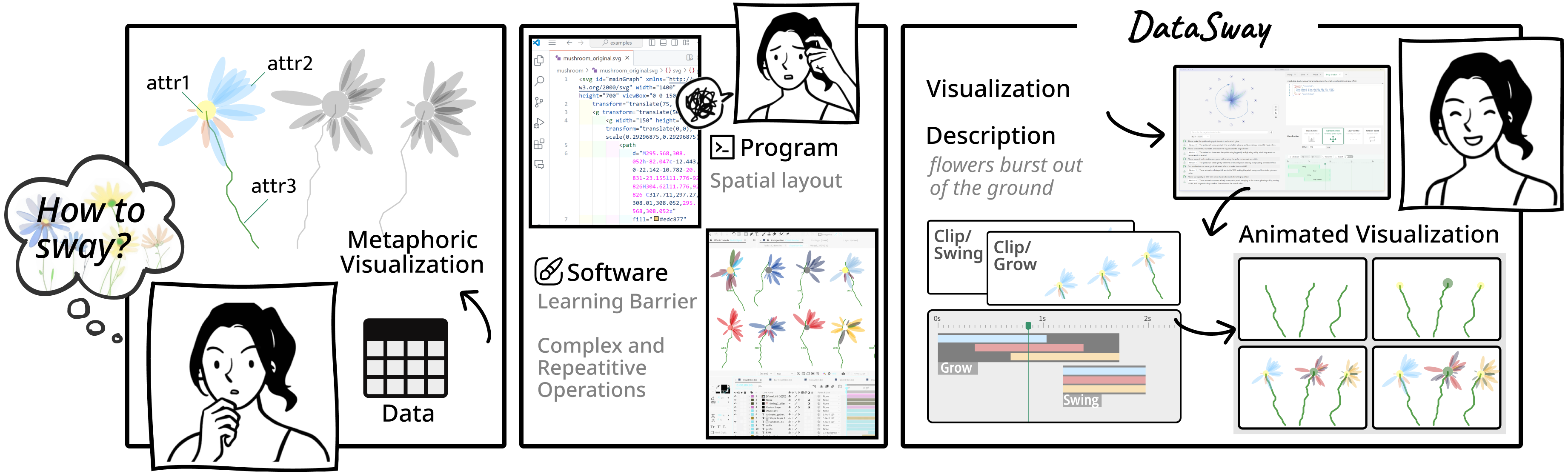}
  \caption{Animating metaphoric visualizations is challenged by coordinating data-driven elements spatially.
  We propose DataSway, a human-AI co-creation tool for vivifying SVG-based metaphoric visualizations through customized animation.
  Based on a visualization reference and textual description, it generates element-wise animation clips and supports fine-grained coordination, including group-wise offsetting and timeline arrangement.}
  \Description{Figure 1 highlights the challenge in animating metaphoric visualization and the proposed solution. The mismatch between human perception and computer-generated animation often hinders the realization of creative vision. To fill this gap, we introduce DataSway, an animation authoring tool that vivifies metaphoric visualizations. Integrated with a VLM-powered conversational assistant, it synthesizes an SVG animation program for atomic data elements for user descriptions and allows fine-grain control in a visual interface.}
  \label{fig:teaser}
\end{teaserfigure}


\maketitle

\input{sections/01_introduction}

\input{sections/02_related_work}

\input{sections/03_formative_study}
\input{sections/04_framework}

\input{sections/05_interface}
\input{sections/07_gallery}

\input{sections/06_implementation}
\input{sections/08_user_study}

\input{sections/09_discussion}
\input{sections/10_conclusion}
\input{meta/ack}
\newpage
\balance

\bibliographystyle{ACM-Reference-Format}
\bibliography{ref}

\appendix
\newpage

\input{sections/11_appendix}

\end{document}

%% file: meta/commands.tex
\usepackage{graphicx}
\usepackage{soul}                       
\usepackage{url}
\usepackage{hyperref}                   
\usepackage{balance}
\usepackage{calc}                       
\usepackage{xspace}                     
\usepackage{multirow}                   
\usepackage{enumitem}                   
\usepackage{listings}
\usepackage{fancybox}
\usepackage{breakurl}
\usepackage{xcolor}
\usepackage{array}

\AtBeginDocument{

}

\setlist{noitemsep,parsep=0pt,partopsep=0pt, leftmargin=10pt} 

\definecolor{lightpink}{RGB}{237,157,202}
\definecolor{lightred}{RGB}{210,121,121}
\definecolor{lightorange}{RGB}{230,170,50}
\definecolor{lightgold}{RGB}{210,194,121}
\definecolor{lightgreen}{RGB}{121,210,121}
\definecolor{lightaqua}{RGB}{121,206,210}
\definecolor{lightblue}{RGB}{121,124,210}
\definecolor{lightpurple}{RGB}{153,102,255}
\definecolor{superlightgreen}{HTML}{E9F4EC}
\definecolor{superlightgreenborder}{HTML}{C8DDCE}
\definecolor{red}{RGB}{178,34,34}
\definecolor{gray}{RGB}{166,166,166}
\definecolor{systemgreen}{RGB}{132,201,214}
\definecolor{baselinepink}{RGB}{132,201,214}

\newcommand{\mybar}[2]{%
  \color{#2}\rule{#1 cm}{6pt}
}

\newlength{\dpcircle}
\newlength{\rcircle}
\newlength{\dcircle}

\newcommand{\ie}{{i.e.,}\xspace}
\newcommand{\eg}{{e.g.,}\xspace}
\newcommand{\ea}{{et~al.}\xspace}

\newcommand{\etc}{{etc.}\xspace}

\newcommand{\repo}{\href{https://github.com/shellywhen/datasway/}{https://github.com/shellywhen/datasway/}\xspace}
\newcommand{\tool}{DataSway\xspace} 
 
\newcommand{\user}{Zoey\xspace}

\newcommand{\rev}[1]{\textcolor{black}{#1}}
\newcommand{\misc}[1]{\textcolor{black}{#1}}

\newcommand{\iq}[1]{``{\it{#1}}''}

\newlength\myheight
\newlength\mydepth
\settototalheight\myheight{Xygp}
\newcommand*\inlinegraphics[1]{%
  \settototalheight\myheight{Xygp}%
  \settodepth\mydepth{Xygp}%
  \raisebox{-\mydepth}{\includegraphics[height=1.2\myheight]{#1}}%
}

\newcommand*\inlineicon[1]{%
  \settototalheight\myheight{Xygp}%
  \settodepth\mydepth{Xygp}%
  \raisebox{-\mydepth}{\includegraphics[height=1.4\myheight]{#1}}%
}

\definecolor{codegreen}{rgb}{0,0.6,0}
\definecolor{codegray}{rgb}{0.5,0.5,0.5}
\definecolor{codepurple}{rgb}{0.58,0,0.82}
\definecolor{backcolour}{rgb}{0.95,0.95,0.92}

\lstdefinestyle{mystyle}{
    backgroundcolor=\color{backcolour},
    commentstyle=\color{codegreen},
    keywordstyle=\color{magenta},
    numberstyle=\tiny\color{codegray},
    stringstyle=\color{codepurple},
    basicstyle=\ttfamily\footnotesize,
    breakatwhitespace=false,
    breaklines=true,
    captionpos=b,
    keepspaces=true,
    numbers=left,
    numbersep=5pt,
    showspaces=false,
    showstringspaces=false,
    showtabs=false,
    tabsize=2
}

\usepackage{listings}
\usepackage{xcolor}

\definecolor{codebg}{rgb}{0.96,0.96,0.96} 
\definecolor{codeborder}{rgb}{0.85,0.85,0.85} 

\lstdefinestyle{prettyinline}{
  backgroundcolor=\color{codebg},     
  basicstyle=\ttfamily\small,         
  breakatwhitespace=false,            
  breaklines=true,                    
  captionpos=b,                       
  commentstyle=\color{gray},          
  keywordstyle=\color{blue},          
  stringstyle=\color{teal},           
  numbers=none,                       
  numberstyle=\tiny\color{codeborder},
  rulecolor=\color{codeborder},       
  frame=single,                       
  framexleftmargin=2pt,               
  framexrightmargin=2pt,              
  framextopmargin=2pt,                
  framexbottommargin=2pt,             
  framesep=2pt,                       
  xleftmargin=3pt,                    
  xrightmargin=3pt,                   
  aboveskip=5pt,                      
  belowskip=3pt,                      
  showstringspaces=false,             
  upquote=true,                       
}

\newcommand{\prop}[1]{%
    \begingroup
  \setlength{\fboxsep}{0pt}%
  \colorbox{gray!5}{%
    \raisebox{0pt}[0.6\height][0.3\height]{%
      \hspace{0.1em}
      \textcolor{darkgray}{\texttt{#1}}%
      \hspace{0.1em}
    }%
  }%
  \endgroup
}

\newcommand{\vmark}[1]{%
    \begingroup
  \setlength{\fboxsep}{0pt}%
  \colorbox{gray!5}{%
    \raisebox{0pt}[0.6\height][0.3\height]{%
      \hspace{0.1em}
      \textcolor{teal}{\texttt{#1}}%
      \hspace{0.1em}
    }%
  }%
  \endgroup
}

\newcommand{\op}[1]{%
    \begingroup
  \setlength{\fboxsep}{0pt}%
  \colorbox{superlightgreen}{%
    \raisebox{0pt}[1.1\height][0.3\height]{%
      \hspace{0.1em}
      \textcolor{black}{#1}%
      \hspace{0.1em}
    }%
  }%
  \endgroup
}

%% file: meta/authors.tex
\author{Liwenhan Xie}
\email{liwenhan.xie@connect.ust.hk}
\orcid{0000-0002-2601-6313}
\affiliation{%
  \institution{The Hong Kong University of Science and Technology}
  \city{Hong Kong SAR}
  \country{China}
}

\author{Jiayi Zhou}
\orcid{0000-0003-4669-4872}
\email{jzhoudp@connect.ust.hk}
\affiliation{%
  \institution{The Hong Kong University of Science and Technology}
  \city{Hong Kong SAR}
  \country{China}
}

\author{Anyi Rao}
\orcid{0000-0003-1004-7753}
\email{anyirao@connect.ust.hk}
\affiliation{%
  \institution{The Hong Kong University of Science and Technology}
  \city{Hong Kong SAR}
    \country{China}
}

\author{Huamin Qu}
\orcid{0000-0002-3344-9694}
\email{huamin@ust.hk}
\affiliation{%
 \institution{The Hong Kong University of Science and Technology}
   \city{Hong Kong SAR}
    \country{China}
}

\author{Xinhuan Shu}
\orcid{0000-0002-9736-4454}
\email{xinhuan.shu@newcastle.ac.uk}
\authornote{Corresponding author}
\affiliation{%
  \institution{Newcastle University}
  \city{Newcastle}
  \country{United Kingdom}
 }

\renewcommand{\shortauthors}{Liwenhan Xie, Jiayi Zhou, Anyi Rao, Huamin Qu, and Xinhuan Shu}

%% file: meta/abs.tex

Animating metaphoric visualizations brings data to life, enhancing the comprehension of abstract data encodings and fostering deeper engagement.
However, creators face significant challenges in designing these animations, such as crafting motions that align semantically with the metaphors,  maintaining faithful data representation during animation, and seamlessly integrating interactivity.
We propose a human-AI co-creation workflow that facilitates creating animations for SVG-based metaphoric visualizations. Users can initially derive animation clips for data elements from vision-language models (VLMs) and subsequently coordinate their timelines based on entity order, attribute values, spatial layout, or randomness. Our design decisions were informed by a formative study with experienced designers (N=8). We further developed a prototype, DataSway, and conducted a user study (N=14) to evaluate its creativity support and usability. A gallery with seven cases demonstrates its capabilities and applications in web-based hypermedia. We conclude with implications for future research on bespoke data visualization animation.

%% file: meta/keywords.tex
\ccsdesc[500]{Human-centered computing~Visualization systems and tools}
\ccsdesc[500]{Human-centered computing~Interactive systems and tools}
\ccsdesc[300]{Human-centered computing~Web-based interaction}

\keywords{Authoring Tools, Motion Graphics, Human-AI Co-creation, Expressive Visualization}

%% file: sections/01_introduction.tex
\section{Introduction}
A {\it metaphoric visualization} leverages familiar imagery or symbols to represent data, which is effective in capturing audience attention~\cite{amini2018hooked}, enhancing retention~\cite{borkin2013makes}, and attaining affective resonance~\cite{lan2023affective, lan2025more}.
One of the famous examples is the \textit{Dear Data} project~\cite{lupi2016dear}.
It encodes data with free-form sketches resembling everyday metaphors like flowers, which turn data into intriguing drawings and vividly speak to the eyes.
In web-based hypermedia, some metaphoric visualizations have moved beyond static representation.
For instance, {\it Four Seasons}~\cite{fourseasons}, 
visualizes performance metrics of each country with a tree swaying cozily in the wind.
The viewers can also interact with the leaves and drill down for country-wise information.
This delicate animation entangles viewers in an immersive experience, enhancing engagement through dynamic motion.

Nevertheless, creating animated metaphoric visualizations is non-trivial.
First of all, translating creative ideas into animation specifications can be time-consuming and skill-demanding~\cite{liu2025logomotion, shen2024dataplaywright}, as it involves precise control over individual data elements and their groups to generate animations that cohere with the metaphor, such as the natural sway of trees.
In addition, when animating a visualization, it is also important to preserve the integrity of data mappings (termed as {\it graphical integrity}~\cite{tufte1983visual}), ensuring that the visual channels encoding data are kept consistent throughout the animation.
Furthermore, as these metaphoric visualizations are typically embedded in interactive web environments, designers must ensure that animations integrate seamlessly with user interactions without disrupting the responsiveness or interpretability.

Despite rich tooling efforts for metaphoric visualization authoring, most works are confined to a static output~\cite{xia2018dataink, liu2018dataillustrator, ying2022MetaGlyph, offenwanger2024datagarden}.
Current visualization animation tools focus on standard chart types and template-based animation~\cite{thompson2020understanding}, which are not readily applicable for meaningful animation design to augment metaphor expressions in bespoke visualizations.
General animation software like After Effects~\cite{aftereffect} assumes a fixed number of graphical elements, which can hardly address data-driven graphics, where the graphical elements can be dynamically rendered due to data transformations via interactions. 
Designers in our interview study see potential in animating visualizations, yet lack handy tools to integrate delicate animation into their works.
\rev{Importantly, we distinguish animation’s role in this context as vivifying the metaphor—using motion to breathe life into the conceptual imagery (e.g., making a tree sway)—rather than adding new data-encoding channels.}

More recently, there have been explorations in VLM-powered motion graphics creation~\cite{liu2025logomotion, tseng2024keyframer, wu2024aniclipart, bourgault2025narrative}, where language-oriented animation synthesis lowers the barrier of implementing ideas with animation primitives.
However, existing works target general motion graphics, lacking control over data-driven elements and visualization-specific constraints on graphical integrity during animation.
Based on the idea of language-oriented animation, we further explore the scope of metaphoric data visualizations, particularly focusing on key aspects of data visualization such as graphical integrity and interactivity.
This is achieved through the careful design of user interaction workflows tailored to the nuances of visualization.

Through a formative study \rev{(\autoref{sec:formative})}, we interviewed experienced designers to learn their perspectives and expectations in design support for animated metaphoric visualizations.
We also analyzed high-quality instances to characterize animation patterns in these artifacts.
Informed by the early insights, \rev{we distill design goals for an animation tool (\autoref{sec:dg}) and propose a workflow to create bespoke animation for metaphoric visualizations (\autoref{sec:framework})}.
Based on an SVG-based static visualization, users may attain element-wise animation clips by describing the target animation to a conversational assistant.
The animation clips are linked to data-bound elements rather than graphical primitives and can be rearranged along a timeline for complex visual effects.
Then, the users can introduce slight variations for each target data group according to a weight value from entity sequences, attribute values, spatial layout, or pure randomness.
\rev{Assigning weights to individual elements within a data group facilitates guided coordination in their sequencing.}
We instantiated the workflow with a prototype authoring tool \rev{(\autoref{sec:tool})} and demonstrated its expressiveness with a hypothetical walkthrough \rev{(\autoref{sec:walkthrough})} and a gallery (\rev{\autoref{sec:gallery})}.
In a user study,
participants generally recognized \tool's usability and usefulness and confirmed its support in design exploration and animation authoring \rev{(\autoref{sec:eval})}.

In general, our work contributes to the following aspects.
\begin{itemize}
      \item A formative study summarizing existing practices, challenges, and tooling expectations on animated metaphoric visualization.
      \item \rev{An animation workflow for metaphoric visualizations. It features VLM-powered keyframe generation to augment metaphoric design and scaffolds the coordination of data-driven elements.}
      \item \tool, a prototype that implements the workflow.
      The open-source code is available at \repo with a gallery of output examples.
      \item A user study evaluating the usability and creativity support of \tool. We further discuss the implications of interface design for bespoke animated visualizations.
\end{itemize}

%% file: sections/02_related_work.tex
\section{Background and Related Work}
\label{sec:rw}
\rev{This section situates our work within the evolving landscape of metaphoric data visualization, animation authoring toolkits for visualizations, and more generic GenAI-assisted graphic design.}
\subsection{Metaphoric Data Visualization}
Despite a historical pursuit for simplistic design~\cite{tufte1983visual}, a growing body of literature has demonstrated the benefits of incorporating metaphors in visualizations for communicative purposes, e.g., improving data comprehension~\cite{woodin2021conceptual, borgo2013glyph}, explaining complex concepts in scientific storytelling~\cite{preim2023survey, pokojna2024language}, and creating affective and engaging experiences~\cite{tkachev2022metaphorical, silva2022field, lan2023affective}.

Pokojna et al.~\cite{pokojna2024language} summarized four types of visual conceptual metaphor use: (1) imagistic metaphors that employ graphic similarity, such as embedding data into thematic images~\cite{coelho2020infomages, gao2025sceneloom}, (2) orientational metaphors that leverage spatial orientation and composition, such as wheels~\cite{alsallakh2012reinventing} for radial layouts, (3) ontological metaphors that map abstract concepts into concrete entities, such as Chernoff faces~\cite{chernoff1973use}, and (4) structural metaphors that relate an entire set of unfamiliar sources to familiar targets.
Following their work, Guo et al.~\cite{guo2026unpacking} proposed a structured design space derived from 2,029 metaphoric infographics to systematically guide the creation of visual metaphors.
There are many authoring toolkits for metaphoric visualizations.
Some support users in integrating metaphoric images in visualizations by repurposing figurative assets~\cite{zhang2020DataQuilt, coelho2020infomages, shi2025piccl}, incorporating sketch-based marks~\cite{romat2019expressive, xia2018dataink}, structure~\cite{offenwanger2023timesplines}, \misc{and their composition~\cite{offenwanger2024datagarden, shi2025comprehensive}}. 
MetaGlyph~\cite{ying2022MetaGlyph} generates glyphs that encode multivariate data with different components of metaphoric images. 

Only a few works have explored animation in metaphoric visualizations.
Lu et al.~\cite{lu2020enhancing} experimented with three types of data-driven animation design to augment static charts, yet they did not consider metaphor consistency. 
Data Pictorial~\cite{zhou2024DataPictorial} deconstructs raster infographics to create animated vector posters with the retrieved data.
DataWink~\cite{xie2025datawink} uses chained VLMs to reverse-engineer SVG visualizations and add animation by updating the underlying D3 program.
Extending this line of research, we practically assume known data, and focus on designing human-VLM collaboration workflows for rapid authoring of meaningful animations for an SVG-based visualization while preserving interactivity.

\subsection{Animation Toolkit for Data Visualization}
One may leverage libraries like D$^3$~\cite{bostock2011d3}, Processing.js~\cite{reas2006processing}, and Anime.js~\cite{anime} to create customized animation, using keyframe-wise or time-based specifications.
However, when writing animation programs, it is not intuitive to coordinate spatial properties of visual elements, which can be critical to metaphor expression.
In addition, lengthy code may hinder careful design exploration and iteration. 
In practice, many designers turn to animation software like After Effects~\cite{aftereffect}.
These tools are effective for creating polished, non-interactive motion graphics but fall short when it comes to dynamic, web-based visualizations---a common form of metaphoric visualizations---that require interactivity and continuous updates based on user input or real-time data.

To lower the barrier for non-programmers, many authoring tools provide interactions and automatic features for users to create animated visualizations freely. 
For example, Data Animator~\cite{thompson2021DataAnimator} adopts a keyframe-based authoring paradigm. 
InfoMotion~\cite{wang2021infomotion} automatically applies entrance animated effects to visual elements in static infographics in a temporal order.
Another stream of these tools focuses on visualization animation in data videos. 
For instance, AutoClips~\cite{shi2021AutoClips} provides an animation library for different data insight types. 
Recent works focus on narration-driven generation of data videos~\cite{wang2024wonderflow, cao2023dataparticles, shen2024dataplaywright} where the textual narration guides the segmentation of animation clips and determines corresponding animated effects.
However, the output animation of these works is mainly preset and based on standard chart templates.
In comparison, we aim to create expressive animations that align with metaphoric semantics, attending to the spatial dynamics of visual elements.
These aspects are not captured in the design space of most visualization-specific graphical interfaces or automatic algorithms.

Valuing the communicating role of animation~\cite{chevalier2016animations},
Shi~\ea~\cite{shi2021communicating} empirically studied how animation can communicate eight narrative strategies, such as emphasis and comparison.
Lan~\ea~\cite{lan2021kineticharts}, designed Kinetichart, a suite of animation templates that project delicate animation with five positive emotions.
Xie~\ea~\cite{xie2023wakey} proposed a technique to transfer the vivid motion of a driving GIF to vector-based text or word clouds, where they showed that the underlying emotion can also be transferred.
Xie~\ea~\cite{xie2023emordle} studied the design of emotionally resonant animated word clouds and proposed an approach to propagate a word-based animation clip to word clouds with randomness or spatial-aware mapping.
Inspired by this approach, we developed an animation framework with more coordination modes and exemplified it as an authoring tool for metaphoric visualizations with a higher abstraction level.

\subsection{GenAI-Assisted 2D Graphics Design}
Our study, although distinctly rooted in animated, metaphorical visualization, shares common ground with general 2D graphic design.
A key intersection lies in the need for precise manipulation of graphical primitives to craft the ultimate visual output.
Apart from a wide adoption for generating or retrieving design assets (\eg~\cite{zhou2024epigraphics, zhou2026collaposer}), GenAI has been applied in various design activities, such as translating text into design primitives~\cite{kim2022stylette}, refining intermediate results in design iteration~\cite{zhou2024understanding}, and synthesizing feedback according to design heuristics~\cite{duan2024generating}.
To combat the ambiguity of textual prompts, some works introduced multimodal representations to surface the design constraints~\cite{masson2023directgpt}.

This work lies between interactive systems affording fine-grain control desired by visual creators and systems featuring AI generation with human optimization in a text-to-design process. 
More recently, Bourgault~\ea~\cite{bourgault2025narrative} introduced AniMate, a hand-drawn animation system that integrates direct manipulation and natural language interaction.
Specific to data visualizations, existing works featuring GenAI support mainly concern static visual encoding design~\cite{chen2024beyond, xiao2023spark, zhou2024epigraphics}.
Due to the complexity of creating comprehensible visuals while ensuring information correctness, these works commonly design human-in-the-loop pipelines allowing intervention.
For instance, AnalogyMate~\cite{chen2024beyond}, an infographics tool for images that draws an analogy to numbers, follows a three-step generation: candidate metaphors synthesis, prompt extension, and visualization generation, which supports the adjustment of intermediate results.

Some works deliberately require LMs to generate parametric animation keyframes for fine-grain control~\cite{tseng2024keyframer, liu2025logomotion, ma2025mover, chen2025dancingboard,shi2026notational}.
For instance, LogoMotion~\cite{liu2025logomotion} presented an automatic pipeline synthesizing animation to display the layered visual elements in a logical sequence.
However, metaphoric visualization differs from these artifacts because it concerns data-driven visual elements and requires high-level coordination for a harmonious scene.

%% file: sections/03_formative_study.tex
\section{Formative Study}
\label{sec:formative}
Given the limited real-world examples of animated metaphoric visualizations and the lack of public discourse on the topic, we conducted a formative study to gather insights from designers with experience in creating metaphoric visualizations.
This enabled us to explore their perspectives on animation practices, challenges, and expectations for authoring support tailored to metaphoric visualizations.
Based on the findings, we formulate the design goals for an animation tool for web-based metaphoric visualizations.

\begin{table*}[!t]
\centering
\caption{Demographics of interviewees in the formative study, with an elaboration on occupation, sex, age, design experience in years (``DE''), visualization experience in years (``VE''), authoring experience with animated visualizations (``AV''), toolkits they have used in designing metaphoric data visualizations or animated graphics, and their self-rated coding expertise.}
\label{tab:interview_demo}
\setlength{\tabcolsep}{4.8pt}

\begin{tabular}{llrrrllll}
\toprule
\textbf{ID} & \textbf{Sex} & \textbf{Age} & \textbf{DE} & \textbf{VE} & \textbf{Design Field} & \textbf{AV} & \textbf{Toolkits} & \textbf{Coding} \\
\midrule
P1 &  Male & 32 & 12 & 8 & UX/UI & No  & D3.js, Processing.js, After Effects, Premiere, Figma & Advanced \\

P2 &  Female & 28 & 10 & 2 & Media art & Yes  & Unity, After Effects & Intermediate\\

P3 &  Male & 29 & 3 & 6 & Graphics design, Fine arts & No & Three.js, Paper.js, After Effects, PowerPoint & Proficient \\

P4 &  Female & 29 & 6 & 2 & Graphics design & No & Illustrator, Figma, Balsamiq Wireframe & Beginner \\

P5  & Female & 27 & 7 & 1 & UX/UI, Graphics design& Yes & D3.js, Sigma.js, After Effects, Procreate, PowerPoint  & Proficient \\

P6 & Female & 26 & 4 & 3  & UX/UI, Graphics design  & Yes& D3.js, P5.js, Three.js, Unity, After Effects, Figma & Advanced\\

P7 & Female & 30 & 7 & 1 & UX/UI, Graphics design & No & D3.js, PhotoShop, Procreate, Figma & Intermediate \\

P8 & N/A & 25 & 3 & 3 & UX/UI & Yes  & D3.js, P5.js, Unity & Advanced\\
\bottomrule
\end{tabular}\\
 {\small \raggedright Sex: ``N/A'' for chose not to disclose. \par}
\end{table*}

\subsection{Set-up}

\paragraph{Participants} 
We recruited eight participants (P1--P8), detailed in \autoref{tab:interview_demo}.
All participants have created motion graphics.
While their average data visualization experience was 3.25 years (SD=2.49, Range 1--8), all participants had experience creating SVG-based metaphoric visualizations, and half had designed or implemented animated visualizations.
This hands-on expertise, combined with their backgrounds in visual design, qualifies them to provide informed perspectives on the challenges of authoring animated metaphoric visualizations.
\rev{More details are available in \autoref{appendix:formative_demographics}}.

\begin{table*}[t]
\caption{A tentative classification of animation design in metaphoric data visualization based on the role of data-driven elements for shaping the metaphor. For each class, we provide an example with the participant's envisioned animation design.}
\label{tab:space}
\setlength{\tabcolsep}{4.8pt}
\centering
\small
\begin{tabular}{p{3.6cm}p{2.7cm}p{3.8cm}p{2.8cm}p{3.1cm}}
\toprule
\multicolumn{1}{c}{\textbf{Unit as Metaphor}} & \multicolumn{1}{c}{\textbf{Glyph as Metaphor}} & \multicolumn{1}{c}{\textbf{Metaphor Descendant}} & \multicolumn{1}{c}{\textbf{Part-of-a-Metaphor}} & \multicolumn{1}{c}{\textbf{Part-of-a-Scenery}} \\
\midrule
 
\parbox[c]{\linewidth}{\centering The data element is metaphoric, and the animation strengthens its meanings.} & 
\parbox[c]{\linewidth}{\centering Some attributes in a metaphorical glyph are animated.} & 
\parbox[c]{\linewidth}{\centering Abstract data elements are embedded in a visual hierarchy, and the animation enhances its metaphoric ascendant.} & 
\parbox[c]{\linewidth}{\centering Data elements compose a strong metaphor with animation.} & 
\parbox[c]{\linewidth}{\centering Multiple metaphorical data elements compose an animated scenery.} \vspace{0.2cm}\\

\parbox[c]{\linewidth}{\centering \includegraphics[height=2.95cm]{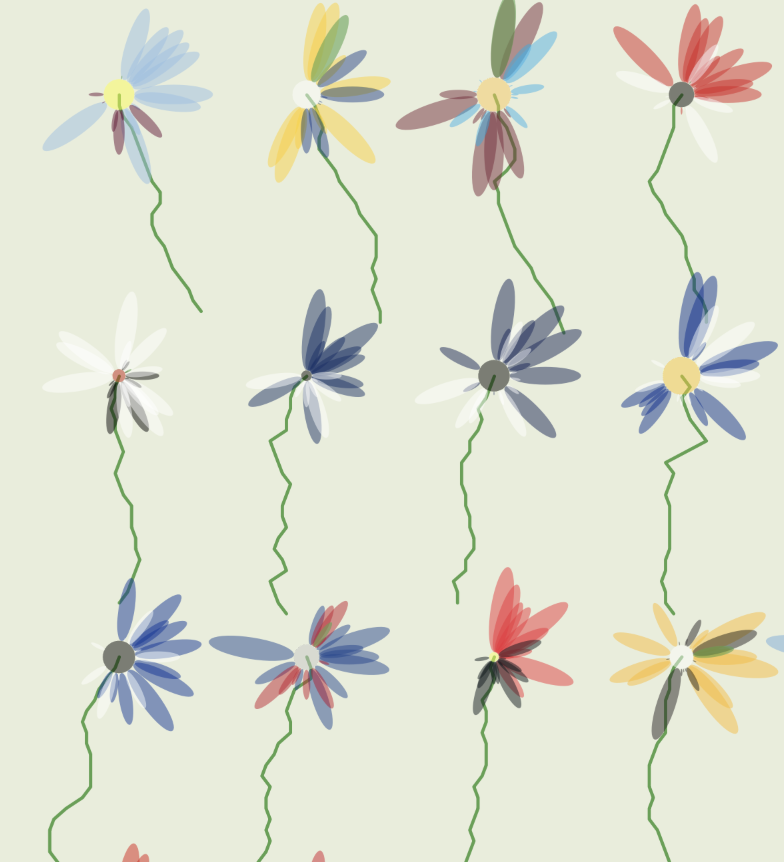}} & 
\parbox[c]{\linewidth}{\centering \includegraphics[height=2.95cm]{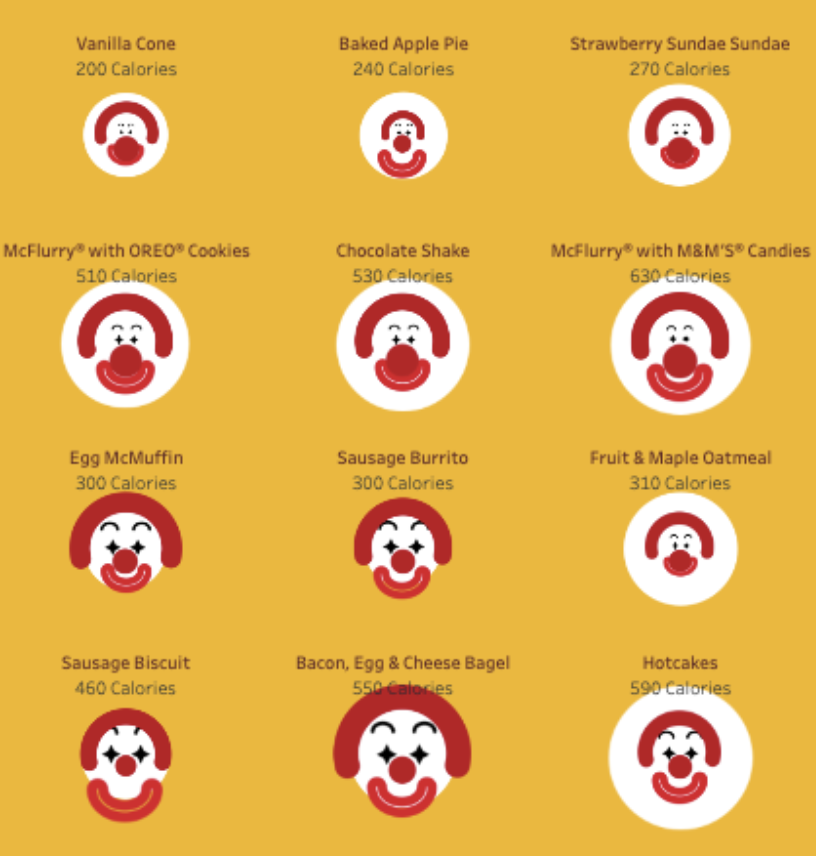}} & 
\parbox[c]{\linewidth}{\centering \includegraphics[height=2.95cm]{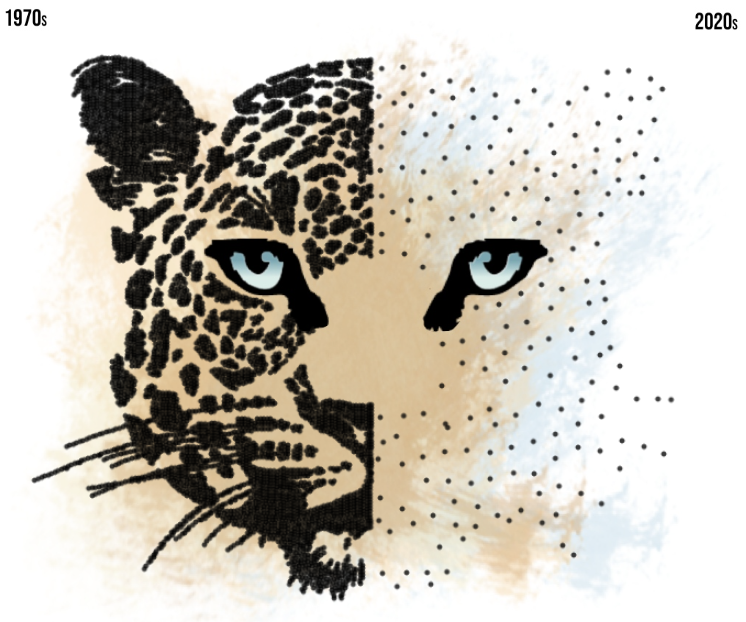}} & 
\parbox[c]{\linewidth}{\centering \includegraphics[height=2.95cm]{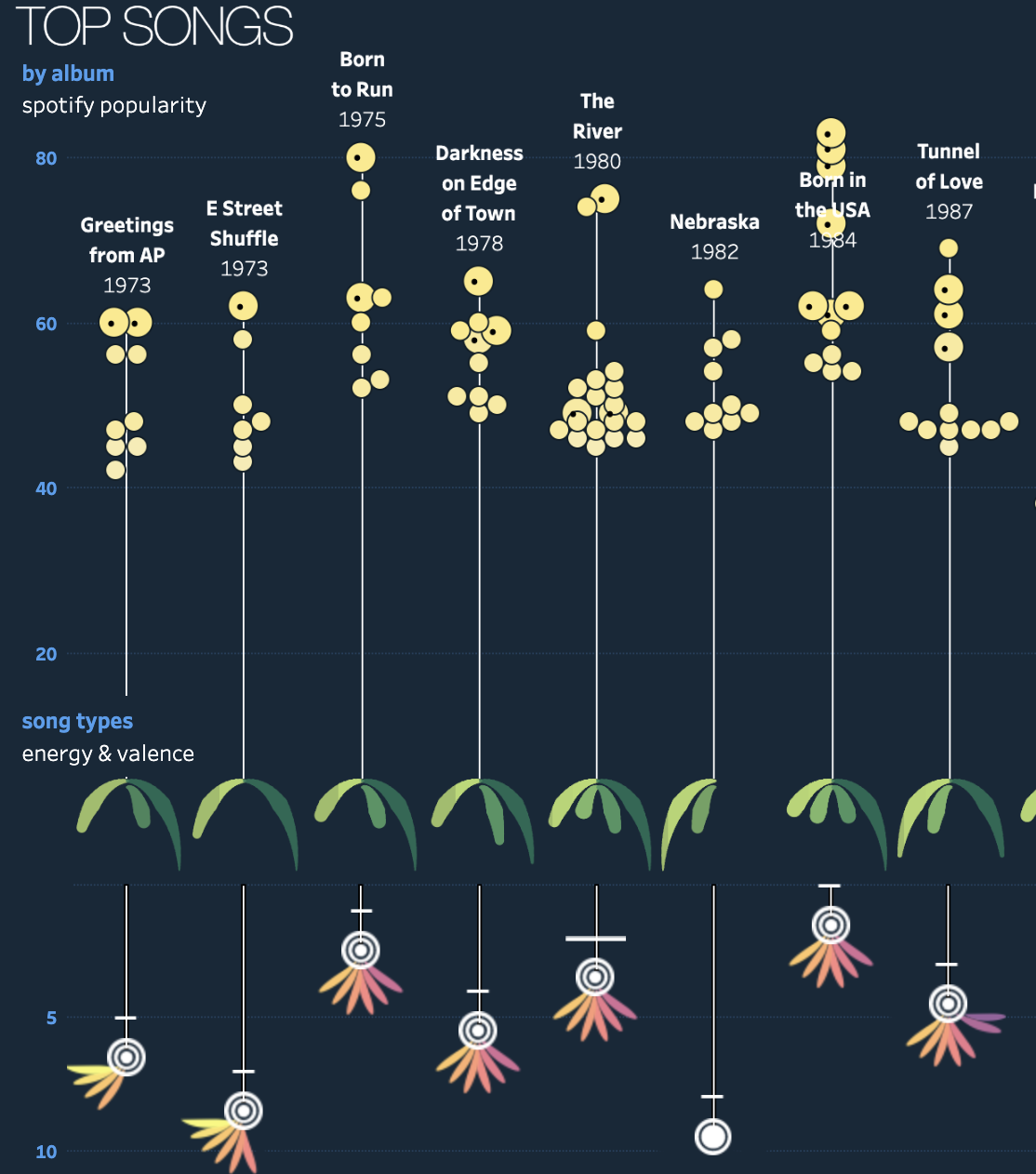}} & 
\parbox[c]{\linewidth}{\centering \includegraphics[height=2.95cm]{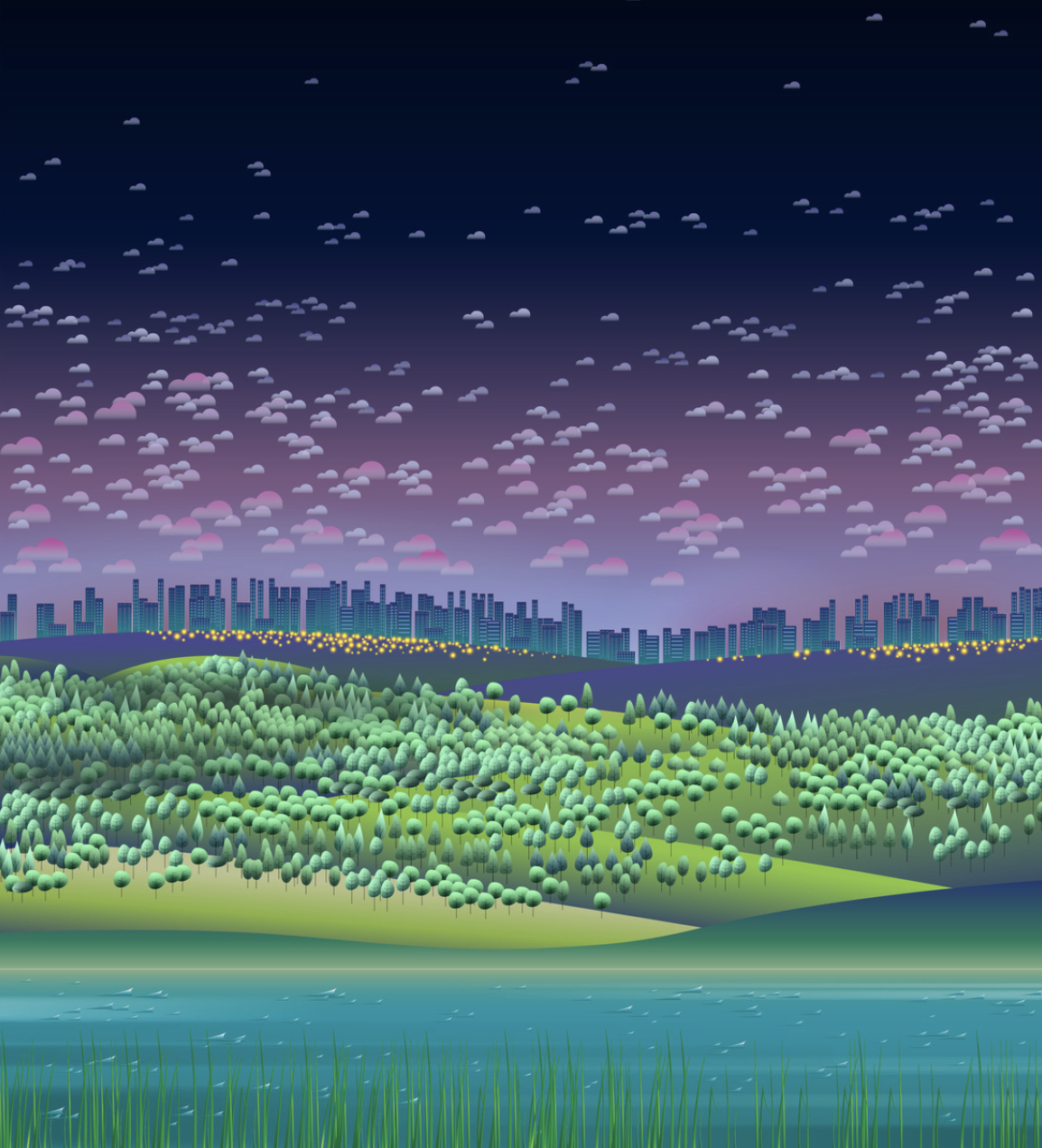}} \vspace{0.15cm}\\

\parbox[c]{\linewidth}{\centering \textit{The petals are glistening.} [\autoref{example:squads}]}& 
\parbox[c]{\linewidth}{\centering \textit{The clown smiles.} [\autoref{example:clown}]} & 
\parbox[c]{\linewidth}{\centering \textit{The leopard roars.} [\autoref{example:leopard}]} & 
\parbox[c]{\linewidth}{\centering \textit{The flowers grow up from the ground.} [\autoref{example:bruce}]} & 
\parbox[c]{\linewidth}{\centering \textit{Clouds are drifting while the trees are swaying.} [\autoref{example:despair}]}\vspace{0.05cm}\\

\bottomrule
\end{tabular}

\Description{Table 2 presents a classification of animation design in metaphoric data visualization. The categories (Unit as Metaphor, Glyph as Metaphor, Metaphor Descendant, Part-of-a-Metaphor, and Part-of-a-Scenery) are arranged as columns. The rows include Explanation, Example Description, and Picture. Each explanation and example description is centered and describes the respective animation design. A placeholder row for pictures is included.}
\end{table*}

\paragraph{Materials} We curated two corpora with 20 animated metaphoric visualizations (Corpus I) and 30 static ones (Corpus II) to facilitate the interview (\autoref{appendix:materials}).
Corpus I illustrates our target output, whereas Corpus II stimulates discussions of potential animation design, aiming to probe the design space of animated metaphoric visualization.
While not aiming for comprehensiveness, we strove to retrieve diverse and representative samples and elicit insights by reviewing entries in various data art venues and the designer forums.
The collected instances cover advanced designs like pictograms, deformed layouts, and glyph-based designs.
A full list is available in an interactive browser at \url{https://datasway.notion.site}.

\paragraph{Procedure} Each interview session consists of a semi-structure interview (25--40 minutes) and a speculative design activity (30--50 minutes).
First, the demographic information is collected.
Then, the interviewer briefly introduces the vision of the project with Corpus I and asks questions following a semi-structured template (see \autoref{appendix:formative_template}).
Next, the participant should pick at least five instances from Corpus II and brainstorm on potential animation designs to enhance its intuitiveness and user engagement.
The interview ends with supplementary comments about previous questions.
The interviews were conducted one-on-one over Zoom.

\paragraph{Analysis} All interview sessions were video-recorded.
The first author took notes for key responses to the semi-structured interview and manually linked them with the original quotes.
Three authors discussed the identified themes and resolved conflicts in the initial interpretation through two group meetings, which lasted an average of one hour.
The findings were organized according to the core research questions of interest.
As this study concentrates on the symbolic role of visual representations in metaphorical visualization, we excluded comments where animation plays a functional role, such as highlighting state changes~\cite{ren2017ChartAccent} or smoothly transiting two visualizations~\cite{kim2021Gemini}.
In analyzing the design patterns of animation, we labeled (1) the role of data-driven elements in contributing to the expression of the metaphor during animation, (2) the group interactions for individual data-driven elements in strengthening the metaphor, and (3) the factors in visual design that significantly contribute to the metaphor expression.

\subsection{Findings}
Here, we summarize the key takeaways from the interview study.
\subsubsection{Design Considerations of Animation in Metaphoric Visualization} Summarizing the self-reflections on relevant experiences and comments on the study materials, we distilled three major design considerations (DC) of animation in metaphoric visualization.

\ul{DC1. Engage the audience with animation design at different times in the data experience.} One prominent strength of animation is its charm in vivifying objects and stimulating emotional resonance (8/8).
During the speculative design activity, participants conceived the animation designs for various interaction events.
This includes a gradual reveal during the visualization's initial loading (8/8), recurrent subtle animation during its display (8/8), customized highlights when focusing on one data entity (6/8), transitions to the visualization of filtered data (3/8), and interactive scaffolding of the encoding structure in a legend (2/8).
Apart from attracting attention, animation can reinforce the static design and set up the affective tone of the data story.
\iq{It matters to make people feel the message instead of interpreting it.} (P2)
In addition, animation may lead to surprise---\iq{A carefully crafted animation is like a little gift. It entices me to keep exploring individual data.} (P4)

\ul{DC2. Align animation with its metaphor.}
Most participants (6/8) recognized that animation is expected to align with its target metaphor.
As metaphoric visualizations commonly employ abstract visual encoding, animation can disambiguate what it resembles by implying its inherent physical properties.
\iq{A stripped circle can be a balloon, but when it bounces, you know it is more likely a ball.} (P5)
As such, animation not only plays a functional role but also facilitates understanding what the metaphor means and informs the underlying context when an audience onboards a data story.
Otherwise, mismatched animation design may weaken the hook for the audience, undermining the design efforts of its static appearance.
\iq{When hovering on a flower leads to dropped shadows, I feel less immersed since it reminds me of a button on a dashboard.} (P3)
Correspondingly, static objects like a wall or rock are expected to stay still (P7).

\ul{DC3. Balance visual complexity in dynamic scenes.}
Despite the identified benefits of animation, some participants (3/8) mentioned its negative impacts in increasing information loads.
In some cases, the static design of a metaphoric visualization (\eg~[\autoref{example:snow}]) can be overly complex, encoding more than three data attributes in different visual channels.
When all the visual components are animated, the audience may be overwhelmed and fail to process and interpret the information.
If animation causes confusion or information overload, it should be removed to simplify the reading experience.
\iq{After all, animation plays a subsidiary role [in enhancing the original metaphor]. If it causes negative impacts, then it should be removed.} (P8)
P1 further noted animation design requirements according to accessibility guidelines\footnote{\url{https://www.w3.org/TR/WCAG21/}}, such as avoiding flashing content and consistent speed between the foreground and background.

\subsubsection{Practice \& Challenges in Authoring Animated Effects in Visualization}
All participants (8/8) have experience in creating motion graphics and static metaphoric visualizations, but only half of the participants (4/8) have created animated metaphoric visualizations.
In addition to working alone and configuring the animation specification from scratch, they may collaborate with a development team.
To illustrate their ideas to a collaborator, they (4/8) commonly describe the target animated effect verbally.
Besides, to convey the nuances in design, some (2/8) may make mid-fidelity mock-ups of the keyframes using graphics tools.
P6 noted the importance of a reference: \iq{Based on a concrete animation example, it is easier to elaborate on what I hope to achieve.}
However, some prominent challenges (C) hindered the participants from implementing animation into their works.

\ul{C1. Limited skill in translating target animated effects into animation specifications.}
While recognizing the value of animation, many participants explained that writing animation programs is mandatory yet difficult for interactive visualizations.
It requires translating the conceived visual effect into concrete numeric configurations on various animation properties.
\iq{This is what I am not trained for. I can only work with templated snippets. It is completely different from setting up the keyframes based on a timeline [in GUI-based tools].} (P7)
This highlights a significant gap in the design handoff process, echoing previous studies~\cite{walny2019data,tseng2024keyframer}, which underscores the need for more designer-friendly tools.

\ul{C2. Complex configuration to coordinate data-driven visual elements.}
Some participants (5/8) resonated that coordinating data-driven elements in animated metaphoric visualizations presents a unique and complex challenge.
Unlike general motion graphics, metaphoric visualizations often contain numerous data points, each represented by a visual metaphor.
These visual metaphors across the scene should maintain spatial relationships while conveying data-driven information.
Therefore, designers must carefully orchestrate the movement, scale, and positioning of numerous metaphoric elements simultaneously, ensuring that their relative positions and interactions make sense both spatially and contextually.
This becomes particularly complex when dealing with elements representing different scales or dimensions of data, as their movements must remain meaningful within the visualization's spatial layout.

\ul{C3. Labor-intensive process to prototype and iterate.}
Some participants (3/8) commented that they were held back by the labor-intensive workflow in prototyping and iterating animation design in metaphoric visualizations.
Additionally, when an error emerges, it could be challenging to identify problematic configurations and correct them correspondingly (2/8).
\iq{Debugging is a headache. If not set up properly, a data element may be broken down into disjoint parts.} (P2)
P5 further pointed out that due to substantial workloads, they usually fixate on an idea without exploring alternatives.
\iq{While I can think of several effects, it is painful and costly to try another animation scheme in the messy code.} (P5)

\subsubsection{Characterizing Potential Animation Design for Metaphoric Visualizations} \label{sec:space}
By classifying the animation of Corpus I and articulated animation on Corpus II, we distilled common patterns of (envisioned) animated metaphoric visualization.
Our analysis focuses on how the target data elements in the animation design relate to the expression of the metaphor, where detailed codes are available in the aforementioned browser.
\rev{Accordingly, we categorize metaphoric animations into five types according to the structural relationship between data elements and the overall metaphor, shown in  \autoref{tab:space}. This spectrum ranges from cases where a single data element itself serves as the metaphor (Unit-as-Metaphor) to complex scenarios where multiple data entities collectively form an animated environment (Part-of-a-Scenery).
This classification reveals that animation does not just add motion; it vivifies the metaphor by either enhancing local details (\eg~Glyph-as-Metaphor) or reinforcing the global narrative (e.g., Metaphor-Descendant).}
We also found that the animated effect described is mostly straightforward and has a short duration (about 1 to 5 seconds) for one take.
This may be because the purpose of the animation is to vivify the underlying metaphor and enchant the audience (DC1).
The animation takes a ``{\it subsidiary role}'' according to P8 (DC3), which is not intended for constructing visual narratives with complex animation design.

\section{Design Goals}
\label{sec:dg}
Drawing from the interview and our experience in authoring animated visualizations, we formulated five design goals (DG) for an animation tool for metaphoric visualizations.

\textbf{DG1. Generate expressive animation from a textual description aligned with static metaphors.} \rev{We are motivated to address the challenge of translating high-level authoring intents of expressive animation into low-level specifications by automatically generating animation and supporting fine-tuning details (C1)}.
An animation specification concerns target elements and motion sequences.
Leveraging state-of-the-art code generation techniques, we hope to convert textual descriptions of users to guide animation generation.
This matches a general expectation that the generated animation can be expressive enough to align with the original, expressive metaphor (DC2).
\misc{Compared with direct-manipulation systems like After Effects~\cite{aftereffect}, a text-driven approach remains more intuitive for users~\cite{tseng2024keyframer} and supports more flexible combinations.}

\textbf{DG2. Enable flexible coordination of data-driven visual elements.}
In a metaphoric visualization, the data encoding scheme maps data entities into various visual channels of abstract graphical elements that compose the metaphoric visual representation.
\misc{To create vivid animation coherent with the static appearance, it is necessary to flexibly coordinate the sequencing and timing of these animation target elements so that they may collectively reinforce the original design and collectively shape the narrative (C2).}
\rev{Specifically, the system needs to support the diverse roles of data elements identified in \autoref{tab:space}.
For instance, to support Part-of-a-Scenery, the system must allow individual elements (like swaying trees) to move in a spatially synchronized manner, thereby forming a harmonious scene.}
Based on an element-wise animation clip, we aim to coordinate the animation timeline flexibly to generate a harmonious animation design for the entire visualization.

\textbf{DG3. Facilitate rapid prototyping for diverse design options.}
In general, we aim to lower the barrier to experimenting with multiple design options, encourage creative exploration, and ultimately lead to more effective animated metaphoric visualizations.
Specifically, we hope to design a low-code workflow for simultaneously rapid prototyping and evaluating multiple design options (C3).
In this way, designers may better leverage their creative strengths to balance various design principles and requirements (DC3).

\textbf{DG4. Seamlessly merge into the natural workflow to author metaphoric visualizations.}
On the one hand, we aim to provide creators with freedom in choosing underlying visualization specifications to fit in various workflows.
On the other hand, we hope to address interactivity critical for visualizations and support advanced editing in creators' natural workflow.
Therefore, we anticipate using SVG-based static visualization, a widely adopted visualization representation, as input and output SVG animation scripts.
This ensures smooth integration into the original interactive visualization without imposing additional workload~\cite{satyanarayan2019critical}, which fits the need to engage viewers with animation throughout their interactions with the visualization (DC1).

\textbf{DG5. Steer animation consistency with data encoding.}
While our goal is to leverage animation to bolster metaphors in data visualization, we are aware of the inherent risks~\cite{tversky2002animation}---unmoderated incorporation of animation could inadvertently lead to misleading or confusing visualizations if not properly aligned with the data encoding.
To mitigate this risk, we aim to incorporate a validation mechanism that would assess the consistency between the animation and the underlying data encoding scheme. 
This validator would guide users to make animation choices that enhance, rather than distort, the represented data, thereby preventing misinformation and promoting responsible design practices.

\begin{figure*}[t]
    \centering
    \includegraphics[width=\linewidth]{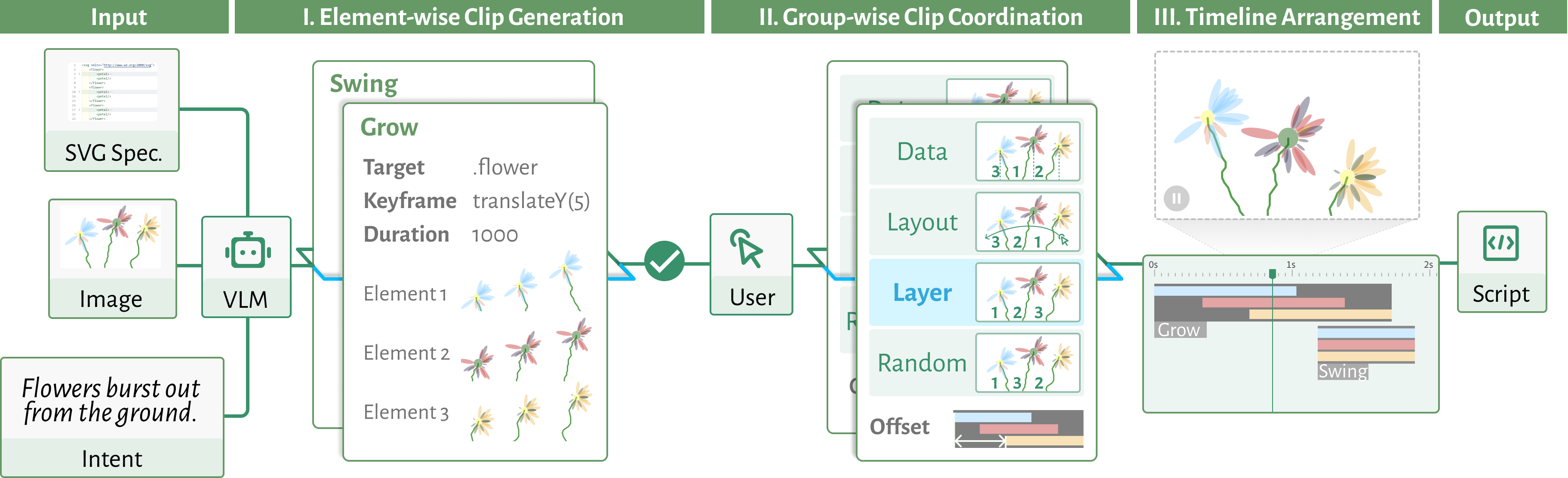}
    \caption{A VLM-powered workflow to create animated metaphoric visualizations. In Stage I, guided by user intent alongside the visualization reference represented in SVG and raster image, a VLM synthesizes element-wise animation clips based on a textual description. \rev{A validator checks the encoding coherence.} In Stage II, user coordinates the animation timing for group-wise data-driven elements with four modes. In Stage III, users may manipulate the clips along a timeline and preview the results for design iteration. The output is an animation script that can be reused for web-based interactive visualizations.}
    \Description{Figure 2 provides an overview of the proposed workflow for animation design in metaphoric visualization. The process begins with the user inputting the visualization and a textual description of the animation into the VLM to generate animation clips. Next, the user can arrange the sequence of these clips based on the SVG structure, spatial layout, data value, or randomly. The animation can be iteratively refined through further conversation and coordination until the desired result is achieved.}
    \label{fig:coordination}
\end{figure*}

%% file: sections/04_framework.tex
\section{Animation Workflow}
\label{sec:framework}
We propose an authoring workflow that centers animation design around data-driven visual elements and their variations and sequencing in the global timeline.
This aligns with a common strategy of practitioners according to the formative study (see \autoref{sec:space}).
As shown in \autoref{fig:coordination}, the workflow comprises three stages: 1) element-wise clip generation, 2) group-wise clip coordination, and 3) global timeline arrangement.
The input is an SVG-based visualization snapshot from a project.
\rev{The output is an animation script that can be integrated into the original visualization project and supports play, pause, and loops in various interaction callbacks (\textbf{DG4}).}

We made two practical assumptions in the input SVG: unified classifiers with meaningful names for data-driven graphical elements and known data values in the elements.
This can be conveniently achieved with a known data mapping scheme, such as deliberately inserting it into the element properties.
Here, a ``data element'' refers to one graphical instance of a data attribute/entity), whereas a ``group'' means a collection of data elements under the same identifier (\eg~same attributes or same values).
For brevity, we use ``animation clip'' and ``clip'' interchangeably in the following.

In Stage I, users describe the target animated effect and attain generated clips (\textbf{DG1}).
These element-wise clips are represented by keyframe-based specifications.
They are uniform for individual visual elements with the same identifier.
Specifically, aligning with standard animation software and libraries, the specification covers primitive graphical properties, including affine transformations like translate, rotate, and scale, as well as other styling attributes.
\misc{Besides, the generated clips are passed to an external sanity check for identifying potential information distortion during the animation (\textbf{DG5})}.
In Stage II, users can tweak the element-wise clip by assigning weights to each instance and introducing animation variations from the weight value.
Informed by the formative study, there are four types of coordination for various expressions of metaphors and relationships between data elements (\textbf{DG2}).
\misc{\textit{Data-centric coordination} relates to the data values bound to the data-driven element.}
\textit{Layout-centric coordination} relates to the spatial layout of the elements.
\textit{Layer-centric coordination} relates to the rendered order of the elements.
\textit{Random-based coordination} means to attach randomness to the elements.
Finally, in Stage III, 
the timeline of each group-wise clip can be adjusted based on the VLM's initial design.

%% file: sections/05_interface.tex
\section{\tool}
\label{sec:tool}
\begin{figure*}
    \centering
    \includegraphics[width=\linewidth]{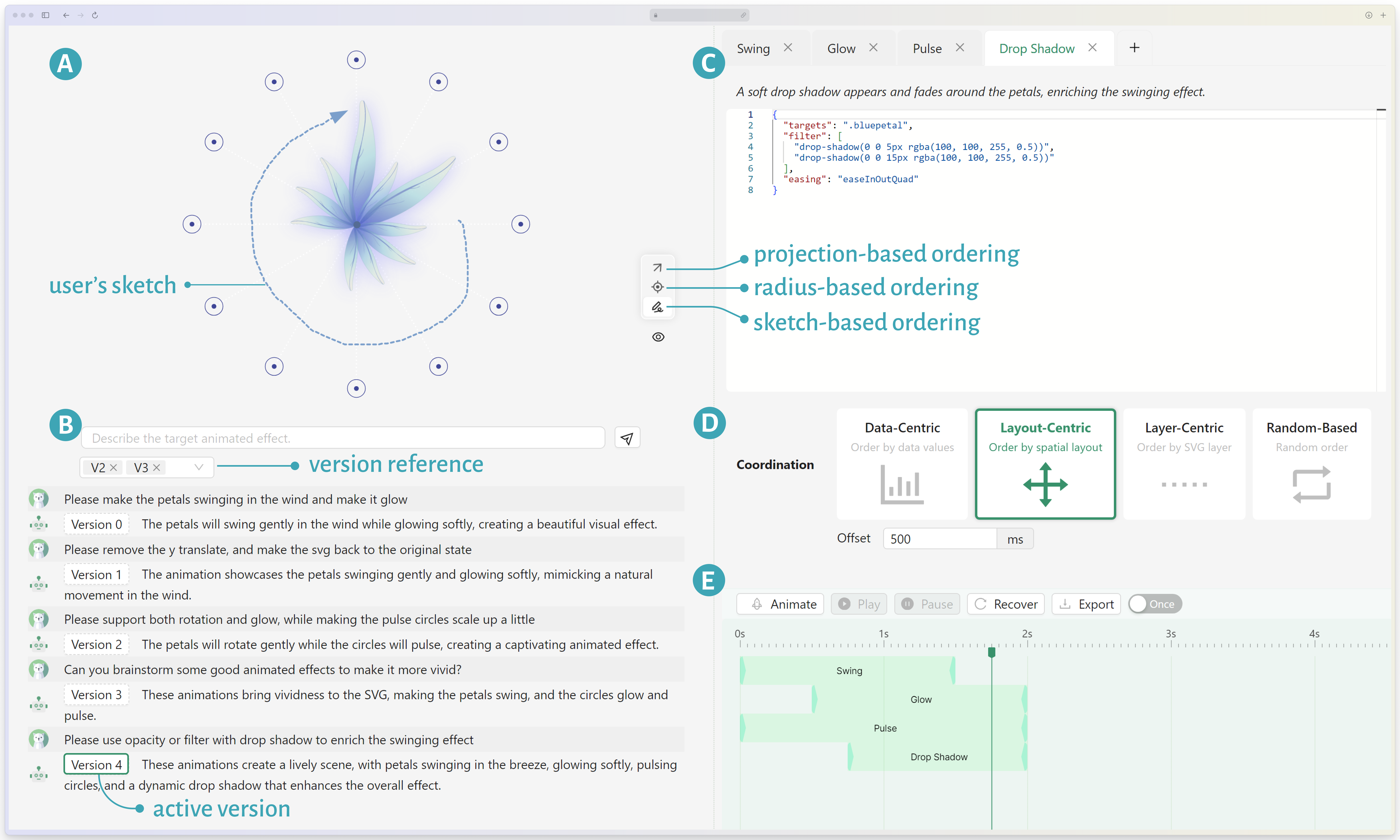}
    \caption{The user interface of \tool. (A) SVG preview panel. Users can also impose layout-centric coordination using direct manipulation. (B) Chatbox panel with a VLM-powered assistant. Users can toggle different animation versions throughout the conversation. (C) Keyframe editor for element-wise animation clips. (D) Group-wise configuration of clips with four types of ordering mechanisms and offset. (E) Timeline panel for configuring the sequencing and duration of the animation scheme.}
    \label{fig:interface}
    \Description{Figure 3 shows the screenshot of the user interface of DataSway. It includes (A) an SVG preview panel on the top left. Users can also impose layout-centric coordination using direct manipulation. (B) A Chatbox panel with a VLM-powered assistant is located on the bottom left. Users can toggle different animation versions within the Chatbox throughout the conversation. (C) A Keyframe editor for element-wise animation clips on the top right. (D) On the middle-right, a Group-wise configuration of clips with four types of ordering mechanisms and offset. (E) On the bottom right, a Timeline panel for configuring the sequencing and duration of the animation scheme.}
\end{figure*}

We present \tool\footnote{The name ``DataSway'' implies that the system supports animating metaphoric data visualizations with vivid visual effects like swaying.}, an authoring tool to design animated metaphoric visualizations.
\autoref{fig:interface} shows a screenshot of the main user interface.
There is another standalone page for configuring inputs.

\subsection{Element-wise Clip Generation}
After uploading the SVG-based reference and style files, the preview panel (\autoref{fig:interface} A) displays the static visualization.
The chat panel (\autoref{fig:interface} B) serves as the entry point and history record for iterative design, where users may leverage a VLM to translate their abstract ideas into basic animation clips. 

\paragraph{Language-guided clip synthesis} The users can describe the intended animation using natural language and derive synthesized clips targeted at specific data elements.
A new sequence of clips is marked as a ``version'', which can be clicked and opened in the right panel for further coordination.
To better communicate the underlying rationales, each clip is associated with a meaningful title and a brief description generated by the assistant.
Besides, users can iterate on some existing versions by choosing the base versions from a dropdown menu and elaborating on how to improve.
Like common conversational LLMs, users can pose questions to communicate with the assistant without generating new clips.
Internally, each user query is wrapped with dedicated prompts for a structured response with a contextual reply, detailed in \autoref{sec:implementation}.

\paragraph{Warning of improper animation design} After new clips are generated, an additional prompt will be sent to the VLM and requests for verification of the compatibility of the animation with the visual encoding.
For instance, animating colors can conflict with an existing encoding that maps categorical values to different colors.
When a suspicious case is found, the associated message raises a warning and elaborates on the reasons as a reminder to the user.

\subsection{Group-wise Clip Coordination}
Initially, when the animation is played, all data elements with the same identifier share identical behavior, as they are controlled by the exact keyframe-based representation of the generated clips  (\autoref{fig:interface} C).
The coordinating panel supports users in manually coordinating these element-wise clips along the time dimension at a group level by imposing an additional element-wise delay through weight calculation and a hyper-parameter---offset (\autoref{fig:interface} D).

\misc{Each element is associated with a weight value normalized to [0, 1].
The offset controls the extent of added delay for an individual element (default as 500 ms) to create dynamic visual effects.}
For instance, if a data element has a weight of $w = 0.1$ and its group-wise clip has a delay of $D = 200$ ms and an offset of $O = 100$ ms, then the starting time of this element would be $t= D + w \cdot O = 210$ ms.
Informed by the formative study, \tool supports various types of weight calculation, covering data value order, positional order, rendering order, and randomness.

\begin{figure}[t]
    \centering
    \includegraphics[width=\columnwidth]{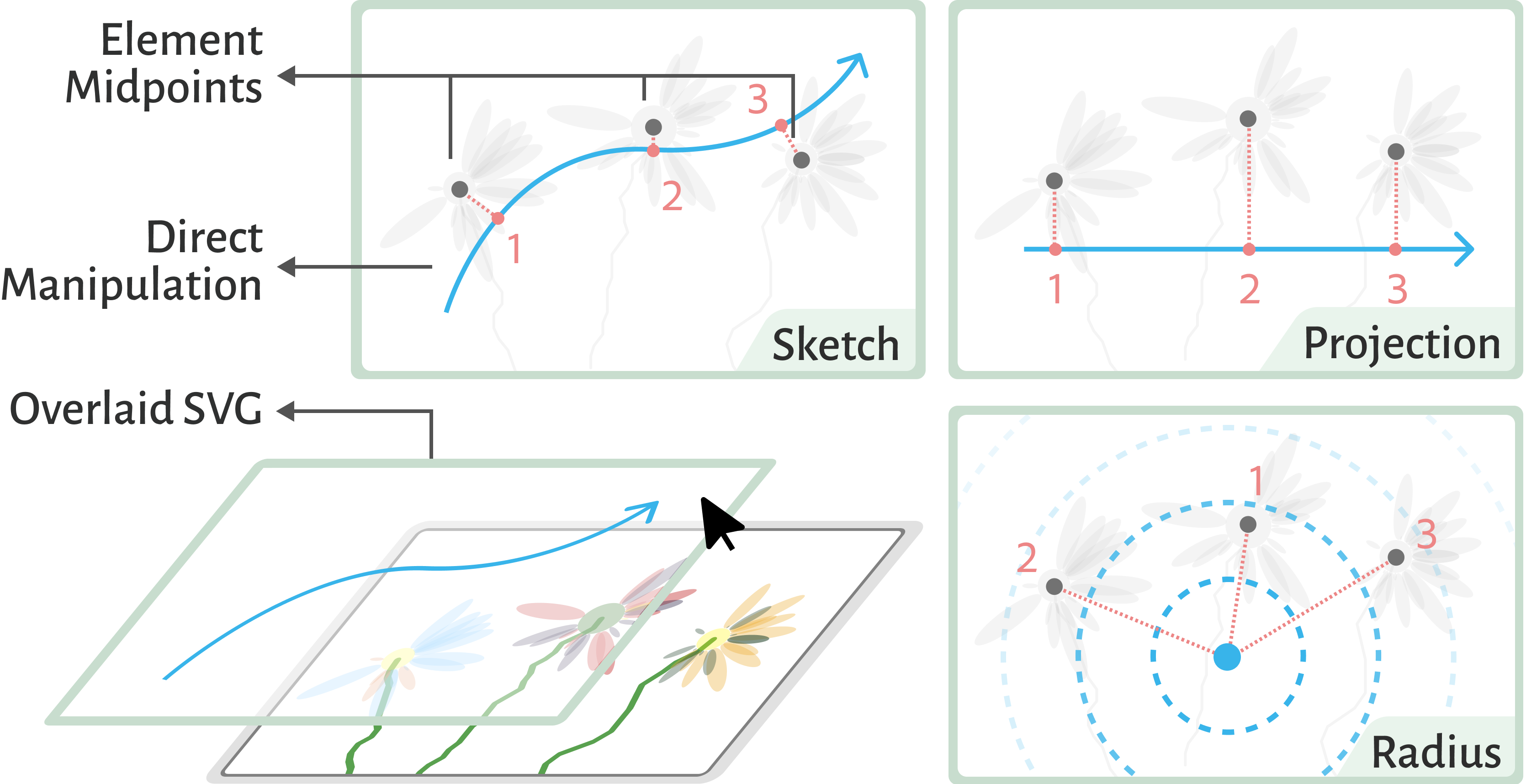}
    \caption{An illustration of spatial coordination supported in \tool, including sketch-based alignment, linear projection, and radius-based ranking. The visual elements are abstracted into a 2D point according to their midpoint.}
    \label{fig:spatial}
    \Description{Figure 4 illustrates four spatial coordination methods supported in DataSway. The sketch-based alignment feature allows users to freely sketch lines on top of the SVG for custom alignment. The linear projection option enables users to draw straight lines for precise spatial arrangement. With radius-based ranking, users can click on the canvas to define a central point and organize elements radially around it. These tools offer flexible control over spatial layout in animation design.}
\end{figure}

\inlinegraphics{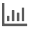} Data-centric coordination---The visual encoding of data is pivotal in metaphoric visualization.
\rev{\tool approximates the quantitative data attribute values by calculating the bounding box's diagonal length of the target visual objects, as metaphoric visualizations commonly map data values into 1D.
This approximation allows accessing runtime data values during interactive sessions.
}
With the data value, one may assign the weight based on a descending or ascending order, and control whether the weights are based on value rank or the numeric value.
\rev{However, future tools may follow a more robust protocol (\eg~dSVG~\cite{ge2020canis}) in data retrieval.}

\inlinegraphics{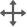} Layout-centric coordination---provides three types of spatial coordination based on the layout of data elements, as depicted in \autoref{fig:spatial}.
In this context, a data element is abstracted into a point on the 2D plane of the SVG canvas.
A \op{radius-based} coordination requires users to click on the canvas to define a center.
The data elements are then assigned weights according to its distance to the center point.
This corresponds to a polar coordinate, commonly used in visualizations.
A \op{projection-based} coordination requires users to draw a straight line, where the weight of an element is proportional to its projection on the line.
The more distant from the starting point, the larger the weight.
Intuitively, this order corresponds to widely adopted Euclidean coordinates in data visualization.
Likewise, a \op{skech-based} coordination requires users to sketch a free-form line on top of the SVG.
A data element will be assigned a weighted value based on the progress of its closest point on the path.
This offers more flexible control, enabling custom paths and other weighting schemes to suit user needs.

\inlinegraphics{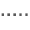} Layer-centric coordination---A layered-centric coordination considers the actual rendering order of data elements in an SVG, as an element declared earlier will usually appear beneath elements declared later, unless specifying \prop{z-index}.
If not considered, there may be occlusion problems in the animation.
Additionally, the rendering order may directly reflect the natural order of the data entities in the dataset, following visualization grammars like D$^3$~\cite{bostock2011d3}.
This can be useful for streaming scenarios and the like with constantly appended data.

\inlinegraphics{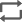} Random-based coordination---The data elements can be randomly assigned a weight value.
This allows for dynamic visual effects where the visual metaphors do not necessarily interact with others, which is one of the common patterns in existing design.

\subsection{Global Timeline Arrangement}
The timeline panel (\autoref{fig:interface} E) surfaces animation clips that compose the current version of animation along a timeline.
To familiarize users, it has a visual design resembling general animation editing tools, including an animation player control bar, a draggable progress bar, and a block-based representation of animation clips in parallel tracks, where each clip takes up a row.
Slightly different from general tools, here, a block represents the group-wise animation clip instead of the individual element-wise animation clip.
Users can resize the block to control its duration and drag to configure its global delay.

\subsection{Interface Design Iteration}
We worked with two long-term collaborators \misc{(P2 \& P5)} in iterating on the interface design.
Based on bi-weekly meetings that lasted about 1 hour for three times, we invited them to try on the prototype system and share their feedback.
At first, the global timeline arrangement was achieved by entering delay and duration values for group-wise clips.
However, the lack of an overview of the temporal relations between clips made it challenging to obtain an intuitive sense of control for the entire sequence.
Therefore, we added the interactive timeline to afford an overview-first, details-on-demand interaction flow (\textbf{DG2}).
In addition, the versioning support in the element-wise clip generation stage was not available in an earlier release of \tool, which overwrote the previous clips and relied on iterative prompting.
However, with an increase in conversational exchange, it becomes necessary to keep track of the history and allow users to revert to a previous version if needed.
Enabling users to precisely refer to specific versions is also essential to guide the assistant to incrementally improve over an existing design instead of generating an entirely new design from scratch (\textbf{DG3}).

\subsection{Implementation}
\label{sec:implementation}
\misc{\tool is a web-based app implemented in the React framework based on TypeScript.
The VLM-based conversational assistant is powered by \texttt{GPT-4o-mini}.
The source code is available at \repo.}

\section{Walkthrough}
\label{sec:walkthrough}
\misc{To illustrate how \tool supports animation creation, we walk through how \user creates animation sequences from \tool and embeds them into a web-based interactive data app.
It is based on {\it OECD Better Life Index}\footnote{A. Pepakayala. \url{https://observablehq.com/@a10k/flower-chart-visual}}, which leverages a floral metaphor to encode development metrics of many countries.}

\misc{After implementing the visualization, \user downloads an SVG and imports it into \tool along with the associated style file. 
\user hopes to strengthen the floral metaphor through an animation depicting growth.
So she types in the chatbox panel: ``{\it Make the flowers grow up}'' and sends.
Then two animated clips are synthesized, marked as Version 1.
The first targets the entire flowers (\vmark{.flower}) with a \prop{translation} along the y-axis and \prop{opacity} from 0 to 1.
The second targets the petal groups (\vmark{.petal}), whereas the \prop{opacity} changes from 0 to 1 with a slight \prop{rotation}.
She plays the sync animation and decides to spread out the animated effect from the center.
So she leverages the radius-based control in the layout-centric coordination and clicks on the SVG center for both clips.
She drags the second clip in the timeline panel to shorten its duration.
Next, she hopes to engage the viewers with some nuanced looped animation.
She types: ``{\it Create a looped animation slowly brightening the petals and darkening them.}
Then, a new animation is synthesized with a warning, saying that \texttt{``\inlineicon{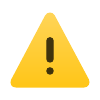}Warning: Color has already encoded metrics. Please make sure that it does not confuse the audience.''}
\user finds it reasonable and asks for more ideas.
Similarly, she makes two other animation sequences.
As illustrated in \autoref{fig:gallery} Case, \user integrates the animation at its first appearance, together with reordering and filtering events.}

%% file: sections/07_gallery.tex
\section{Gallery}
\label{sec:gallery}
\begin{figure*}[t]
    \centering
    \includegraphics[width=\linewidth]{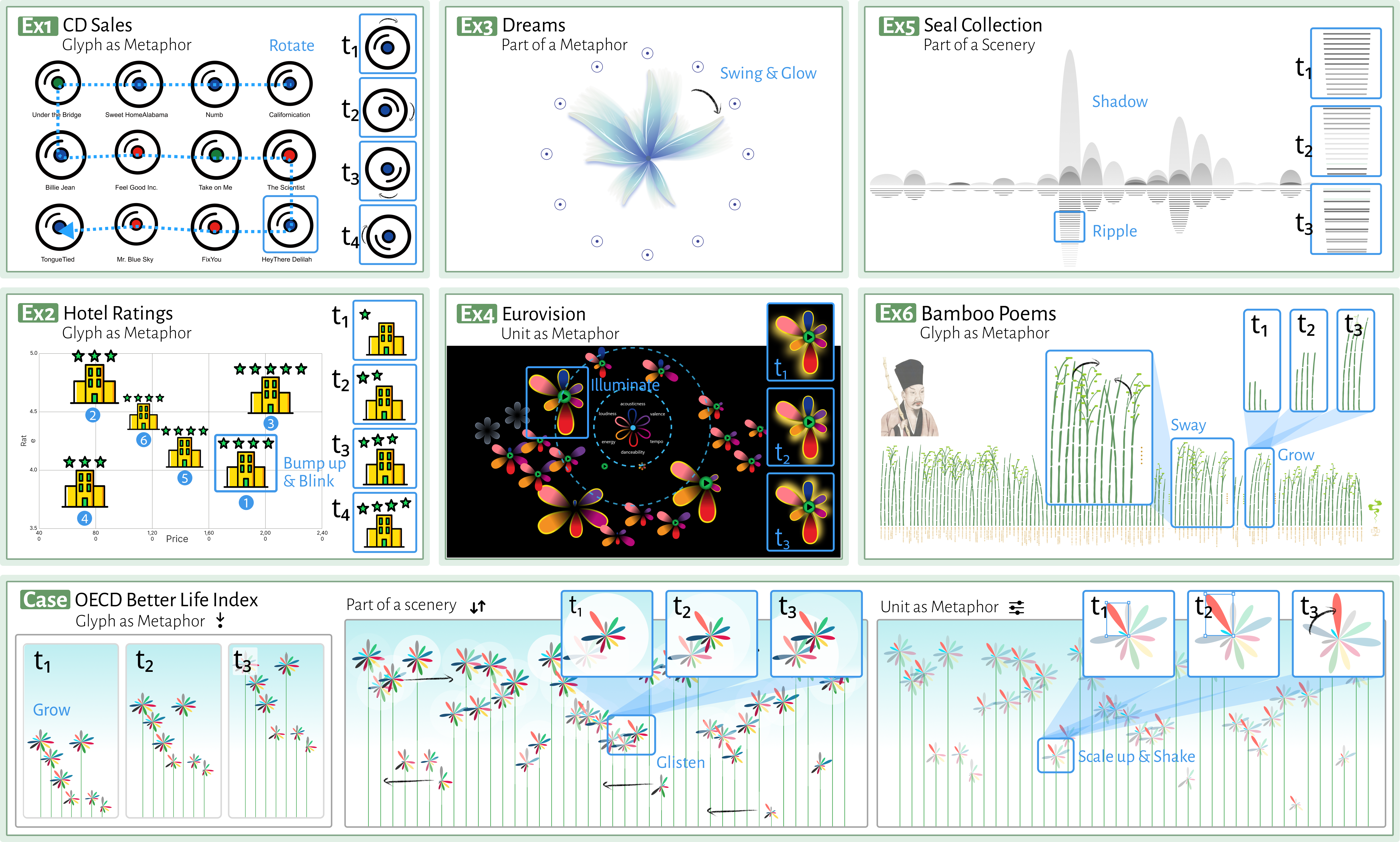}
    \caption{A gallery of animated metaphoric visualizations made by \tool.}
    \label{fig:gallery}
    \Description{Figure 4 demonstrates a gallery of animated metaphoric visualizations, including examples 1 to 6 and the case study made by DataSway.}
\end{figure*}

We demonstrate the expressiveness of our tool with a gallery of seven animated examples (see \autoref{fig:gallery}) created through \tool.
Ex 6 and Case are sourced from user study participants and animated by them in open-ended tasks; others are static metaphoric visualizations collected online and animated by the authors.

\paragraph{\textnormal{\textbf{Ex1.} Glyph as Metaphor}---CD Sales \textnormal{\cite{ying2022MetaGlyph}}} This work leverages a disk-like glyph design to encode three data attributes: color--CD types, size--prize, and shadow--sales.
The grid layout has no data encoding.
Based on an element-wise clip that rotates the CD, the CDs are coordinated by an external sketch.
Therefore, they start to rotate slowly, following the sinuous order \inlinegraphics{figures/icons/spatial.pdf}.

\paragraph{\textnormal{\textbf{Ex2.} Glyph as Metaphor}---Hotel Ratings \textnormal{\cite{ying2022MetaGlyph}}} This chart uses a hotel metaphor for a scatterplot showing information about hotels: number of stars--ranking, x position--price, y position--rate, and glyph size--number of rooms.
The hotels are animated according to their size order \inlinegraphics{figures/icons/data.pdf} from large to small, with the stars bumping up and blinking at the top.

\paragraph{\textnormal{\textbf{Ex3.} Part-of-a-Metaphor}---Dreams \textnormal{[\autoref{example:dreams}]}} This visualization has a circular layout where the petals of an imaginary blossom show the number of dream subjects grouped in one of twelve categories.
When animated, the petals swing clockwise following a sketch-based coordination \inlinegraphics{figures/icons/spatial.pdf} and glow with purple lights.

\paragraph{\textnormal{\textbf{Ex4.} Unit as Metaphor}---Eurovision \textnormal{[\autoref{example:eurovision}]}}
The floral glyphs show songs of Eurovision, with each petal encoding musical metrics like tempo and loudness.
We strengthen the petal's metaphor through an illuminating effect synced along the timeline and played in loops.

\paragraph{\textnormal{\textbf{Ex5.} Part-of-a-Scenery}---Seal Collection} This work leverages a shan shui (hills and water) metaphor to show the collection of 20 artworks.
The hills show the distribution of seal collections in four dynasties.
The ripple corresponds to the seals, represented by a tick in the abstract shadow.
We make the ticks of a ripple randomly \inlinegraphics{figures/icons/random.pdf} move up and down in a looped animation while changing the opacity of the mountains \inlinegraphics{figures/icons/spatial.pdf} to suggest sunlight.

\paragraph{\textnormal{\textbf{Ex6.} Glyph as Metaphor}---Bamboo Poems} This visualization uses poems as a lens to illustrate the life of a famous poet.
Each bamboo represents a bamboo poem.
The x-axis represents the poet's lifespan, while the y-axis depicts the poet's career.
If there are leaves on top of a bamboo, it means that the poem expresses positive emotions.
We create an animation sequence in which the bamboos grow from left to right \inlinegraphics{figures/icons/spatial.pdf} and gently sway.

\paragraph{\textnormal{\textbf{Case}}: OECD Better Life Index}
As introduced in the hypothetical walkthrough, the final product of the OECD Better Life Index contains three animation designs.
The growing animation (glyph as metaphor) starts when the viewer scrolls to the trigger point.
If a viewer reorders the flowers (by country names or by index values), the hidden hallow behind the flower will start glistening in a random order (part of a scenery), suggesting chaos during repositioning.
There is an external control panel that viewers can use to reweight the contribution of each factor and compare by filtering particular data dimensions.
When specific metrics are highlighted, the corresponding petals will scale up and shake (unit as metaphor).

%% file: sections/08_user_study.tex
\section{User Study}
\label{sec:eval}

We conducted an in-lab user study with 14 participants to evaluate the proposed framework and the authoring experience with \tool. We aimed to address the following research questions.

\begin{enumerate}[leftmargin=24pt]
\renewcommand{\labelenumi}{\textbf{RQ\theenumi.}}
    \item Does \tool effectively aid in designing creative metaphoric animations?  (\textbf{DG1})
    \item What strategies do different users adopt with \tool? (\textbf{DG2})
    \item Can \tool support rapid prototyping for animated meta\-pho\-ric visualizations, and what glitches occur? (\textbf{DG3})
    \item What are the perceived pros and cons of using \tool, and how well does it fit in the user's natural workflow? (\textbf{DG4})
\end{enumerate}

\subsection{Participants}
We recruited 14 participants (denoted as U1--U14) through social media and word-of-mouth for people interested in creating animated metaphoric visualizations.
There are 11 females and 3 males, aged 21--28 (Med=25.5, M=24.9, SD=2.45).
They have an average of 3.27 (SD=1.33) years of experience in visualization design and 3.33 (SD=2.96) years in animation design.
Two participants have no animation experience, ten are familiar with video editing tools, and six are familiar with animation programming.
All participants have viewed metaphoric visualization.
In particular, seven have created metaphoric visualizations, and four have created animated ones.
Four participants from the formative study \misc{(P2, P3, P5, and P6)} have agreed to join the user study, now identified as U5, U9, U10, and U12.
We included both expert users and general users in the evaluation process to ensure our animation authoring tool is user-friendly and effective across a range of skill levels.
Given the low entry barrier in \tool, we believe it is essential to capture the experiences of amateur users to identify usability issues and validate the tool's relevance in real-world applications.

\subsection{Protocol}
\indent \paragraph{Tasks}
There are three tasks in a user study session.
The first two are close-ended tasks, where participants should replicate a given animated effect shown in GIFs.
We designed the target animation to cover ontological (glyph as a metaphor) and structural (part-of-a metaphor) metaphors common in our corpora.
Task A is based on Ex1 and requires the rotation of the disks from top to bottom.
Task B is based on Ex3, where participants are asked to make the entire plant swing while glowing.
Task C is open-ended.
Participants should animate one metaphoric visualization from the provided examples to strengthen the metaphor(s) while explaining rationales.

\paragraph{Baseline \& Apparatus}
We replaced the coordination panel and the timeline view with an executable code editor as the baseline system, namely Baseline (referred to as ``chat-only interface'' in the study to mitigate priming effects).
The Baseline resembles conversational programming, a viable approach leveraging nowadays VLM-powered assistants like ChatGPT~\cite{openai2024chatgpt}.
Correspondingly, the underlying VLM was prompted to synthesize free-form animation code instead of structured keyframes as in \tool.
\tool and Baseline were deployed as a web app.
Participants joined the study in person or through Zoom and finished their tasks.

\paragraph{Procedure}
We opted for a counterbalanced within-subjects design to mitigate the learning effect.
Before the study, we obtained signed consent forms and collected demographic information.
The formal study starts with a presentation on our vision of animated metaphoric visualization, and the facilitator walks the participant through \tool and Baseline using an example and responds to any questions (5--10 min).
The participants should then work on Task A, B, and C sequentially.
Each task is limited to 12 minutes, determined through 2 pilot studies.
The participant should try both \tool and Baseline in one task, where the order depends on a randomly assigned group,~\ie Baseline-first or \tool-first.
The study ends with a questionnaire (5 min) and a semi-structured interview (10--15 min).
Thus, a session lasts about 1 hour.
Following the protocol, three of the authors conducted the user study in a one-on-one manner.
Participants were compensated with around \$16 of coupons in local currency.

\subsection{Measures \& Analysis}
We adopted the general Creativity Support Index (CSI)~\cite{cherry2014quantifying} to compare the authoring support in Baseline and \tool, as there is a lack of instruments that measure the authoring support of expressive animation, and authoring effectiveness is an important aspect in creative activities.
We also developed a questionnaire adapted from the System Usability Scale (SUS) based on a 7-point Likert scale to evaluate the usability of \tool (\autoref{appendix:questionnaire}).
In analyzing the system learnability, we dropped the data of the two designers involved in the iterative design phase to ensure fairness.
The participants' interaction activities were automatically logged in Baseline and \tool, including the textual prompts, keyframe modification, coordination details, and timeline arrangement.
Their answers in the post-study interview were summarized in a structured list based on the research questions (\autoref{appendix:post_interview}).

\subsection{Results}
Here, we report our observations in the user study, questionnaire results, and participant feedback in the interview.

\subsubsection{Observations}
\paragraph{Task Performance}
As shown in \autoref{tab:replication}, in close-ended tasks (Task A \& Task B), participants using \tool generally succeeded more in organizing and animating tasks compared to those using the Baseline approach, which struggled with task complexity and required manual input.
In the open-ended task (Task C), many participants leveraged the coordination panel in \tool and created more satisfying results.
\begin{table}
\centering
\setlength{\tabcolsep}{6pt}
  \setlength{\aboverulesep}{0.5pt}
\setlength{\belowrulesep}{0.5pt}
\caption{The number of successful participants (\%) in the replication study (N=14) for clip generation \& coordination.}
\label{tab:replication}
\begin{tabular}{@{}l|cc|cc@{}}
\toprule
 & \multicolumn{2}{c|}{\textbf{Clip Generation}} & \multicolumn{2}{c}{\textbf{Clip Coordination}}\\
 \cline{2-5}
 & Baseline & DataSway & Baseline & DataSway\\
\midrule
\textbf{Task A} & 14 (100.00\%) & 14 (100.00\%) & 2 (14.29\%) & 11 (78.57\%) \\
\textbf{Task B} & 13 (92.86\%) & 13 (92.86\%) & 2 (14.29\%) & 10 (71.43\%) \\
\bottomrule
\end{tabular}
\end{table}

\begin{table*}[!t]
\centering
\caption{The Creativity Support Index (CSI) results comparing Baseline and \tool (N=14).}
\setlength{\tabcolsep}{2.95pt}
  \setlength{\aboverulesep}{0.5pt}
\setlength{\belowrulesep}{0.5pt}
\begin{tabular}{l|r|rr|rrrr}
\toprule
                     & \multicolumn{1}{l}{}             & \multicolumn{2}{|c}{\textbf{Mean Factor Score (SD)}}                   & \multicolumn{4}{|c}{\textbf{Mean Weighted Score (SD)}} \\  \cline{2-8}
                     & \textbf{Mean Factor Count (SD)} & \multicolumn{1}{c}{Baseline} & \multicolumn{1}{c|}{DataSway} & \multicolumn{2}{c}{Baseline}  & \multicolumn{2}{c}{DataSway} \\
\midrule
\textbf{Exploration}          & 4.43 (0.76)                      & 11.43 (4.96)                 & 14.93 (2.81)                 &\mybar{1.45588235294}{baselinepink}& 49.50 (20.56)               & \mybar{1.93029411765}{systemgreen}  & 65.64 (15.26)                \\
\textbf{Expressiveness}       & 3.36 (0.84)                      & 10.64 (4.72)                 & 14.14 (3.23)                 &\mybar{1.03794117647}{baselinepink}& 35.29 (18.22)                 &\mybar{1.41382352941}{systemgreen}& 48.07 (17.66)                \\
\textbf{Results worth effort} & 3.14 (1.46)                      & 11.43 (4.42)                 & 14.36 (2.82)                 &\mybar{1.02323529412}{baselinepink}& 34.79 (21.58)                 &\mybar{1.29823529412}{systemgreen}& 44.14 (23.17)                \\
\textbf{Enjoyment }           & 2.07 (1.33)                      & 10.86 (4.29)                 & 13.93 (3.25)                 &\mybar{0.999584545077}{baselinepink}& 24.36 (24.07)                 &\mybar{0.834117647059}{systemgreen}& 28.36 (19.75)                \\
\textbf{Immersion    }        & 1.21 (0.89)                      & 9.14 (4.20)                  & 11.29 (2.64)                 &\mybar{0.308823529412}{baselinepink}& 10.50 (9.41)                  &\mybar{0.403235294118}{systemgreen}& 13.71 (11.66)                \\
\textbf{Collaboration}        & 0.00 (0.00)                      & 9.14 (4.37)                  & 13.43 (2.71)                 &\mybar{0}{baselinepink}& 0.00 (0.00)                   &\mybar{0}{systemgreen}& 0.00 (0.00)                  \\ \hline
\multicolumn{4}{l}{\textbf{CSI Value}}                                                     &\mybar{1.51382352941}{baselinepink} &51.47 (19.43)                &\mybar{1.96}{systemgreen}  & 66.64 (12.25)               \\
\bottomrule
\end{tabular}
\label{tab:csi}
\Description{Table 2 presents the detailed results of the Creativity Support Index (CSI) comparison between Baseline and DataSway. It includes the average factor score, average weighted score, and average factor count across six categories: exploration, expressiveness, results worth effort, enjoyment, immersion, and collaboration. The CSI value for \tool is 66.64 (SD = 12.24), higher than the Baseline score (M = 51.48, SD = 19.41).}
\end{table*}

$\diamond$ \ul{Close-ended Tasks.} Success in closed-ended tasks was roughly measured by the presence of target behavior due to difficulty replicating exact animations.
In Task A, SVG elements were not in a grid-like order.
The baseline hardly synthesized correct delays due to the incompetence of VLMs' reasoning.
Many tried detailing the layout but failed, losing patience or confidence and ending early (5/14).
Only two completed it by manually inputting element IDs.
With \tool, 11/14 accurately assigned orders using projection-based layout-centric coordination.
The rest (3/14) didn't fully grasp \tool's mechanism, indicated order through prompting, and failed.
In Task B, 13/14 attained a generally aligned animation sequence with clockwise rotation and opacity/shadow changes within two prompts in both conditions.
However, the target swaying started from the left petal.
In Baseline, the same two from Task A achieved this by specifying element IDs.
With \tool, ten replicated it through sketch-based layout-centric coordination.



$\diamond$ \ul{Open-ended Task.} In Task C, participants adopted various strategies to create animation effects.
Participants employed three primary strategies in creating animated metaphoric visualizations.
The first is to write detailed descriptions specifying the target elements, animated effects, and time sequencing in one prompt.
The second is to use minimum expressions (such as a target and a verb) for the animation, then incrementally refine it by supplementing clarifications and selecting satisfying version references.
The last is to chat with VLMs in the first place for diverse ideas (\eg~``{\it How can I animate it?}''), then drill down on a design and gradually refine it.
In elaborating the animated effects with natural language, the participants have used animation terms (\eg~``followed by a pulse effect''), specific values (\eg~``{\it rotate 5 degrees}''), and rhetorical devices (\eg~``\textit{dance with joy}'', ``\textit{crawl like a spider}'', and ``{\it flutter gracefully}'').
On average, for one static metaphoric visualization, participants initiated 7.24 (SD=2.19) chats with Baseline and 5.83 (SD=3.16) chats with \tool.
In \tool, participants spent the most time trying different coordination schemes and timeline arrangements.
Only 5 participants tried to modify the keyframes, whereas the rest talked to VLMs entirely.
In addition, we did not observe cases where participants expressed animation intents that violated the data encoding scheme, thereby not triggering the validator. 
\rev{The absence of triggers suggests that for common metaphoric animations, users’ intuition often aligns with graphical integrity, but it does not diminish the validator's role as a safeguard for more complex or unconventional design explorations.}

\paragraph{VLM Failures}

Baseline produced 253/264 scripts and \tool 195/198 sequences from user prompts, with all output grammatically correct or schema-compliant.
However, 1.19\% of Baseline outputs invoked non-existent library functions.
\tool did not have this problem as we required structured output of element-wise clips.
In both conditions, VLMs struggled to match user requirements reliably.
A prevailing failure is reasoning the spatial relationship of SVG elements based on its XML and image representations.
Another common problem relates to hallucinations that produce invalid keyframe values, \eg~using a nonexistent filter.


\subsubsection{Quantitative Ratings}
The CSI value for \tool is 66.64 (SD=12.25), which is higher than Baseline (M=51.47, SD=19.43).
Specifically, 85\% of participants (12/14) reported higher ratings of creativity support in \tool.
Using a \textsc{Wilcoxon Signed-Rank Test}, there is a statistically significant difference between the two conditions (\textsc{W}=16.0, p=0.010, one-tailed).
A detailed summary is shown in \autoref{tab:csi}.
Generally, in the task of animating static metaphoric visualizations, the participants value design exploration, expressiveness, and results worth the effort.
And \tool performs best in supporting design exploration.

\begin{figure}[t]
\includegraphics[width=\columnwidth]{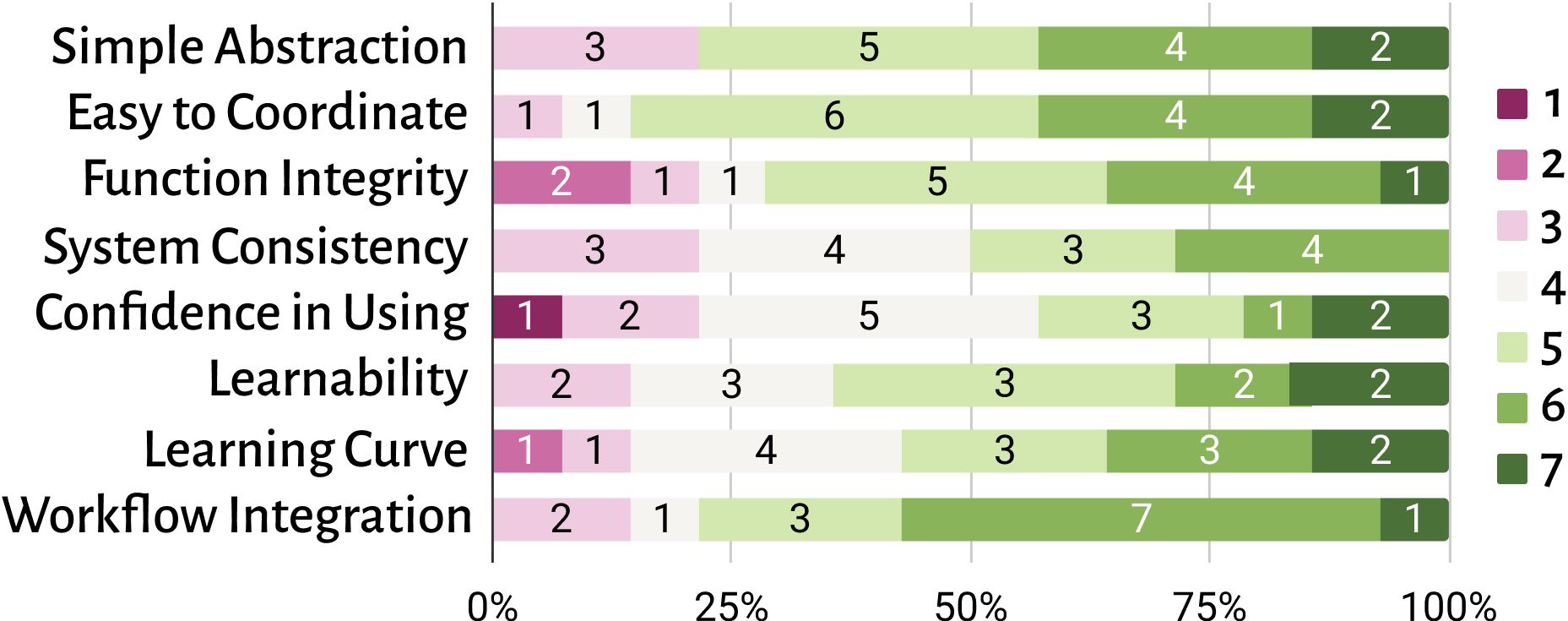}
\caption{Usability ratings on \tool based on a 7-point Likert scale. 1: Highly negative; 4: Neutral; 7: Highly positive.}
\label{fig:questionnaire}
\Description{Figure 6 features a bar chart illustrating usability ratings for the system DataSway, assessed using a 7-point Likert scale, where 1 signifies "highly negative" and 7 denotes "highly positive." The chart includes eight criteria: Simple Abstraction, Easy to Coordinate, Function Integrity, System Consistency, Confidence in Using, Learnability, Learning Curve, and Workflow Integration. Each criterion is represented by a horizontal bar, with varying shades of color indicating the distribution of user responses. The usability ratings for DataSway are broken down as follows: 
"Simple Abstraction" receives three 3 points, five 4 points, four 5 points, and two 6 points. 
"Easy to Coordinate" receives one 1 point, one 3 point, six 4 points, four 5 points, and two 6 points. 
"Function Integrity" receives two 2 points, one 3 points, one 4 points, five 5 points, four 6 points, and one 7 points.
"System Consistency" receives three 3 points, four 4 points, four 5 points, and four 6 points.
"Confidence in Using" receives one 1 point, two 3 points, five 4 points, three 5 points, one 6 points, and two 7 points.
"Learnability" receives two 3 points, three 4 points, five 5 points, two 6 points, and two 7 points.
"Learning Curve" receives one 2 points, one 3 points, four 4 points, three 5 points, three 6 points, and two 7 points.
Finally, "Workflow Integration" receives two 3 points, one 4 points, three 5 points, seven 6 points, and one 7 points.}
\end{figure}

In the usability rating (see \autoref{fig:questionnaire}), the participants are generally satisfied with the animation abstraction, coordination interactions, function integrity, system consistency, system learnability, low learning curve, and workflow integration, with more than half (N$\ge$7) participants showing agreement.
However, participants showed a lack of confidence in the system, with 35.7\% (5/14) rated neutral and 21.4\% (3/14) negative.

\subsubsection{General Impressions} We studied the user experience (UX) between two conditions and summarizing the perceived usefulness and glitches in \tool.

$\diamond$ \ul{UX difference between conditions.}
Compared with Baseline, most participants (10/14) had an increased sense of control in \tool.
U13 was strongly in favor of \tool, commenting that \tool scaffolds the complex authoring process in animation creation for data visualizations---\iq{It makes more sense to decompose animation for fine-grain controls instead of relying on AIs to take care of every detail.}
U2 referred to the coordination procedure as \iq{a more natural staggering function for visualizations} compared with ad-hoc scripts generated in Baseline, and admired its usefulness unseen in off-the-shelf tools.
Some participants with higher programming skills preferred Baseline for its one-page declaration of animation that can be easily modified (U11, U12) and full potential for any animation design (U11). However, as commented by U4---\iq{You cannot count on conversation to control everything.}

$\diamond$ \ul{Perceived usefulness of \tool.}
When being asked for favorite parts of \tool, participants were generally positive about \tool in rapid prototyping and design exploration, especially with the NL-guided clip generation (10/14), versioning support (7/14), group-wise coordination (12/14), and global timeline arrangement (5/14).
P5 appreciated being able to attain usable animation scripts with \tool, explaining that it enables fast realization of rough ideas, whereas \iq{I gave up many ideas thinking of the implementation burden [in the past]}.
Among the four available modes, participants were most excited about the layout-centric coordination (10/14).
\iq{It works like magic and aligns with visual thinking.} (U1)
U7 thought that \tool covered most animation needs for metaphoric visualizations, as \iq{a more complex animation may cause readability issues.}
Participants thought \tool was suited for amateur scenarios with lower requirements on design precision, including personal data storytelling (3/14), engaging data presentation (2/14), and ``jam sessions'' in early project planning (U10).
They appreciated the conversational interaction that led to readily reusable code (10/14).
Most participants (13/14) agreed with the input-output formats and found it fit their workflow.

\subsubsection{\misc{User Challenges in Using \tool}} Several factors account for frustration in using \tool, detailed as follows.

$\diamond$ \ul{The gulf of envisioning.}
As with other prompt-based tools, participants were challenged by {\it the gulf of envisioning}~\cite{subramonyam2024bridging} when interacting with the VLMs, which is reflected in the lack of confidence according to the usability rating.
First, it takes time to gauge the capacity of VLMs and potential design space spanning from the animation framework (8/14).
Second, the performance of VLMs is unpredictable and dependent on the concrete prompts and the conversation history.
Sometimes, wording matters even if the semantics are similar.
For instance, after several trials, U7 accomplished Task B by articulating ``sway'' instead of ``swing''.
Last, though \tool provides short titles and descriptions for the generated clips, some users (3/14) still found it overwhelming to make sense of the generated results.

$\diamond$ \ul{Unfamiliar combination of code editing and video editing.}
Though the participants generally agreed with \tool's workflow (11/14), they had mixed opinions on the tool's interface combining a text editor and a multi-track timeline panel.
Some participants with advanced programming skills (3/14) embraced the text editor, explaining that this allowed them to directly tweak the keyframes and reuse the clips in other applications.
Conversely, several participants with limited coding experience (5/14) suggested replacing the code editor with a standard GUI.
\iq{It's odd to see such low-level code! I only hope to focus on the visual part,} said U13, who used to collaborate with developers.
Participants with both skills generally adapted to this design fast (7/14).

$\diamond$ \ul{Usability issues toward high-fidelity output.}
While participants generally recognize the increased control in \tool compared with Baseline, some (5/14) expressed the need for more controls toward high-fidelity outputs.
U5 requested character motion with trajectory control and advanced rigging, commenting that \iq{[In data storytelling,] data could be personalized and embodied with character movements.}
U10 commented on an inherent limitation of keyframe-based specification: \iq{Sometimes the keyframe values are dependent on relative position to other elements. It is best to conduct separate the calculation instead of wrapping up in the animation clips.}
U10 and U4 hoped to access templates as well, as \iq{Prebuilt modules reduce the effort to create everything from scratch} (U4).

\subsubsection{Opportunities of Future Improvement.}
Participants generally wished for a more fluid interaction that supports in-situ selection of targets directly on the SVG (6/14).
\iq{I would like to click instead of typing in the class name.} (U14)
For prompting, future systems may consider visualizing key structures of SVG and the data mapping scheme to remind users (4/14).
Some participants wished for better guidance in constructing the prompts, such as embedding a cookbook (U6) or supporting automatic refinement (U3, U11).
U1 envisioned \tool as a development plugin directly accessing the visualization for integration testing.
U9 and U12 indicated a need for complex geometry calculation and physical simulation in the synthesized animation for more realistic effects.

%% file: sections/09_discussion.tex
\section{Discussion}
\label{sec:discussion}
Here, we discuss lessons learned, reflect on the limitations of this study, and suggest future research opportunities.

\subsection{Lessons Learned}
\misc{Our exploration reveals that animating metaphoric visualizations is a complex negotiation between creative expression and data integrity.
The following lessons highlight the dual role of animation in enriching communication, the technical hurdles GenAI faces in balancing fluidity with accuracy, and how the {\it intent first, nuances on demand} paradigm provides a robust framework for managing these challenges in human-AI co-creation.}

\paragraph{\misc{Is animation important for metaphoric visualization?}}
Despite the prevalence of charts using abstract visual elements, designers have explored metaphoric data encodings to enhance emotional resonance and information retention~\cite{lan2023affective, snyder2026crossing}.
In our formative study, only a few real-life examples were found, suggesting a minor role for animation.
However, interviewees expressed positive attitudes toward adopting animation.
Designers value animation for visual communication and often envision animated elements to enliven visualizations.
The challenge lies in coordinating data-driven elements to achieve harmonious visual dynamics, which hinders the exploration of this potential.
Our tool was designed to translate creative visions into animation specifications that integrate seamlessly into real-life visualization applications.
Through our user study, we gained confidence in enabling rapid prototyping and design exploration for animated metaphoric visualizations.
Participants acknowledged the enriching impact of animation and showed keen interest in incorporating more of it into their work.

\paragraph{\misc{What challenges GenAI in animating metaphoric visualization?}}
Drawing from our experience, several inherent characteristics in metaphoric visualization make meaningful animation a challenging task for GenAI.
First, the visual elements in an interactive visualization are data-driven, meaning that some visual properties of the animation target, including position, shape, color, etc., are subject to data changes.
The animation should adapt to data changes and ensure graphical integrity, taking various constraints simultaneously.
Second, visualizations embed information and require readability and accessibility.
Animation should not introduce misinformation.
Third, as seen in the user study, a reasonable animation scheme concerns the correct identification of target elements, proper configuration of the behaviors, and time-based orchestration, which spans a large design space and raises challenges.
The last challenge relates to data scalability, where a list of data entities may exceed the context window of LMs or lead to a long processing time.
In our framework, the VLM is responsible for the demanding and tedious task of configuring low-level animation keyframes.
Human creators, however, are responsible for finding proper ways to vivify metaphors at an abstract level and orchestrating clips meaningfully.
The use of a popular animation library well-mastered by commercial VLMs contributes to the mostly reliable outputs in \tool.

\paragraph{\rev{``Intent first, nuances on demand'' paradigm in GenAI-powered design.}}
\rev{\tool follows the design philosophy of {\it intent first, nuances on demand}''~\cite{xie2026intent} and presents a special case study of GenAI-powered design in animated metaphoric visualizations.
The two-stage workflow directly addresses the {\it gulf of envisioning}~\cite{subramonyam2024bridging} by allowing creators to first establish their high-level expressive goals through natural language--leveraging the VLM’s ability to synthesize complex animation specifications from abstract metaphors, and then offer space for professional precision with structured scaffolds like the coordination panels and keyframe editors.
The future of human-GenAI co-creation may lie in fully autonomous generation, but in systems that treat GenAI as a generator of plausible materials while reserving the role of orchestrator for the human designer.
By internalizing domain-specific constraints (\eg~ data encoding consistency) as automated guardrails, tools can empower creators to explore diverse design spaces while maintaining professional standards of accuracy and coherence.
}

\subsection{Limitations and Future Works}
\misc{For expressive data storytelling, future efforts must address three key dimensions: the development of evaluation metrics, the integration of multimodal controls, and the expansion of the design space toward more complex, structurally-aware motion graphics.}

\paragraph{Investigate effective metrics and guidelines for visualization animation}
DataSway enables rapid authoring of bespoke animations for metaphoric visualizations.
However, we lack effective metrics and guidelines to rigorously evaluate the quality of the generated animation.
This is inherently challenging as assessing nuanced animation can be subjective and requires an understanding of the context. 
Our initial efforts focus on ensuring the fidelity of data representation through an automatic validator, which helps maintain consistency between the animation and the underlying encodings. 
On the other hand, our animation is built around communicative visualizations, where we seek to assess higher-level attributes, such as how well the animation fosters user engagement or conveys abstract concepts. 
Despite the valuable empirical insights into the expressiveness and engagement of these animations~\cite{shi2021communicating, Shu21Gif}, they remain subjective and ad-hoc. 
\rev{We expect to move beyond empirical observations toward computational, objective evaluations of aesthetic quality, animation appropriateness, and engagement with audience profiles, leveraging multimodal models. 
This shift extends the call for developing quality metrics for visualizations~\cite{behrisch2018quality} with a particular emphasis on the animation subdomain.}

\paragraph{Combating ambiguity \& expressiveness of natural language input with multimodal controls in animation authoring}
First, while textual descriptions are valued for their intuitive expression of animation intentions, they inevitably introduce ambiguity, especially when specifying targets. 
For instance, terms like ``flowers'' and ``petals'' can both be interpreted as referring to petals, leading to confusion. 
LLMs frequently struggle to resolve such distinctions with precision. 
Second, textual prompts alone are insufficient to accurately and expressively describe animations. 
Our focus on metaphoric visualizations also poses interesting open questions for capturing the structure of elements in SVG, e.g., row- or column-wise layouts and hierarchies~\cite{xie2025datawink, wu2025layerpeeler}, which may be partially addressed through alternative input methods; for example, participants appreciated our system's group-wise coordination features.  
Future systems should further incorporate multimodal interactions beyond textual prompts, such as sketch~\cite{bourgault2025narrative, shi2026notational}, direct manipulation~\cite{zhang2023editing},
and reference~\cite{xie2023wakey}, to support flexible intent expression.
\misc{Furthermore, as our current observations are grounded in short-term sessions, future research should conduct longitudinal studies to identify evolving human-AI collaboration paradigms.
For instance, investigating how a persistent versioning history of half-baked outputs can serve as a reflective tool, enabling creators to backtrack, branch out creative goals, and refine their intent over time~\cite{rawn2023quickpose}.}

\paragraph{Towards expressive animation for metaphoric visualization}
We contribute to a larger effort to explore the design space for animation in visualizations. 
\tool currently supports coordinating element-wise animation clips along the timeline.
However, the potential extends beyond simple temporal sequencing.
First, in addition to a group-wise control of the element-wise delay, animation variation could be introduced by mapping data to keyframe parameters~\cite{lu2020enhancing, xie2025datawink}.
Second, our framework does not address the potential visual hierarchy of data elements.
For instance, in the Metaphor-Descendant class (\eg~[\autoref{example:leopard}]), vivifying the metaphor requires awareness of the global structure in the animation specification.
\rev{Future work may investigate hybrid approaches like integrating physical models~\cite{liu2024spatial, wang2024versatile} or advanced motion graphics tactics~\cite{kazi2014draco, willett2017secondary}, while addressing graphical integrity~\cite{li2025chartgalaxy} and learned representation of natural motions~\cite{li2024GenerativeImageDynamics}.
We envision future authoring systems achieving more profound and exquisite animation for metaphorical visualizations beyond current practice.}



%% file: sections/10_conclusion.tex
\section{Conclusion}
\label{sec:conclusion}
This study probes an understudied topic in animation authoring for metaphoric visualizations in everyday data communication.
Through content analysis of available artifacts (N=50) and interview studies (N=8), we gained a preliminary understanding of the user requirements.
We propose a workflow featuring human-AI co-creation for vivifying SVG-based metaphoric visualization through customized animation. The workflow encompasses VLM-powered element-wise clip generation and group-wise clip coordination.
\tool instantiated the workflow with versioning control in conversation, GUI-based animation coordination in data element offsetting, and timeline arrangement.
As demonstrated in a gallery and case study, \tool affords fast prototyping and design exploration.
A user study (N=14) further evaluated its effectiveness and usefulness, where participants generally appreciated its support in design exploration and rapid prototyping with granular controls.
We regard this work as a starting point toward creating expressive visualization animation powered by AI and call for more scholarly attention to this area.



%% file: meta/ack.tex
\begin{acks}
The authors would like to thank Fanny Chevalier, Rubaiat Habib Kazi, Li-Yi Wei, and Ronghuan Wu for their valuable feedback.

\end{acks}

%% file: sections/11_appendix.tex
\section{Interview Study}
\label{appendix:formative}

\subsection{Participant Recruitment \& Demographics}
\label{appendix:formative_demographics}
We adopted a convenience sampling strategy in the recruitment, aiming to ensure the quality of feedback.
We first invited 5 data artists and designers from our own social network.
Additionally, we reached out to 3 award-winners of a national data art competition (ChinaVis Art Program).
We prioritized candidates with rich experience in communicative visualization design, especially in motion graphics and metaphoric visual encodings.

Please refer to \autoref{tab:interview_demo} for demographic details of interviewees in the formative study.
Among the participants were 5 designers, 2 artists, and 1 PhD student in visualization with a background in graphic design.
They were from different institutes.
We asked them to self-report their coding skills as ``Beginner'' (have not written a visualization program), ``Intermediate'' (can write a visualization program with external help), ``Proficient'' (capable of writing interactive visualization programs and have more than five relevant projects), and ``Advanced'' (have a deep understanding of coding and can perform full-stack development in data-driven applications).

\subsection{Materials}
\label{appendix:materials}
The following are the two corpora with examples of dynamic (N=20) and static metaphoric visualizations (N=30) shown to the participants.
An interactive browser could be found on \url{http://datasway.notion.site}, with manual codes on the metaphoric design.

The corpora were curated based on our review of long-listed entries in the \textit{Information is Beautiful Award}\footnote{\url{https://www.informationisbeautifulawards.com/}}, portfolios of award-winning visualization designers, artworks from the \textit{IEEE VIS Conference Art Program (VISAP)}, popular works from \textit{Tableau Public}\footnote{\url{https://public.tableau.com/app/discover}}.
For Corpus I, we included all instances we found that employ animation techniques in web-based data-driven graphics.
For Corpus II, we kept a representative design with strategies including pictograms, deformed layouts, glyph-based designs, \etc
In addition, to attain diversity in the small number of samples and reduce repetitive patterns, we performed open coding and removed similar items with the same labels.
Specifically, we labeled (1) the role of data-driven elements in contributing to the expression of the metaphor during animation, (2) the group interactions for individual data-driven elements in strengthening the metaphor, and (3) the factors in visual design that significantly contribute to the metaphor expression.
\subsubsection{Corpus I: Animated Metaphoric Visualizations}
\newcommand{\supp}[5]{
    \item {\small#1. #2. {\it #3} #4  \url{#5}}}
\begin{enumerate}[left=8pt]
    \renewcommand{\labelenumi}{[{\small M\arabic{enumi}}]}

\supp{Liuhuaying Yang}{2023}{Four Season: How has the world progressed in recent years?}{}{http://www.go4trees.com/four-seasons/}

\supp{Junlin Zhu}{2020}{Flowing boundary: Data-driven interactive narrative visualization about Covid-19 in China.}{}{https://vimeo.com/506974751}

\supp{Valentina D'Efilippo and Lucia Kocincova}{2018}{MeToomentum: Trending Seeds}{}{http://metoomentum.com/trending.html}

\supp{Federica Fragapane and Paolo Corti}{2020}{The shapes of our dreams.}{Google News Lab.}{https://the-shape-of-dreams.com/\#/chapter2?section=explore}\label{example:dreams}

\supp{Krist Wongsuphasawat}{2017}{Winter is here: Revisit the most discussed moment for each \#GoT character.}{Twitter Inc.}{https://twitter-interactive.vercel.app/winter-is-here/}\label{example:snow}

\supp{Charlotte Qin}{2019}{HeartBees.}{IEEE VIS Arts Program.}{https://www.qintheory.studio/design/heartbees}

\supp{Jin Wu, Weiyi Cai, and Simon Scarr}{2018}{Oil spilled at sea: Putting the Sanchi disaster into perspective.}{Reuters Graphics.}{https://fingfx.thomsonreuters.com/gfx/rngs/OIL-SPILLS/010060SL1GQ/index.html}

\supp{Valentina D’Efilippo and Nicolas Pigelet}{2014}{Poppy Field.}{}{https://www.poppyfield.org}

\supp{Min Lu, Noa Fish, Shuaiqi Wang, Joel Lanir, Daniel Cohen-Or, and Hui Huang}{2020}{Enhancing static charts with data-driven animations.}{IEEE Transactions on Visualization and Computer Graphics.}{https://vizgroup.github.io/activateviz}

\supp{Bon Adriel Aseniero, Sheelagh Carpendale, George Fitzmaurice, Justin Matejka}{2022}{SkyGlyphs: Reflections on the design of a delightful visualization.}{IEEE VIS Arts Program.}{https://www.research.autodesk.com/publications/skyglyphs}

\supp{Jan Wächter, Tania Boa, Mark Hintz, Ilya Boyandin, and Benjamin Wiederkehr}{2015}{Galaxy of covers.}{Interactive Things.}{https://galaxy-of-covers.interactivethings.io}

\supp{Eleanor Lutz}{2017}{An animated chart of 42 North American butterflies.}{Tabletop Whale.}{https://tabletopwhale.com/2014/08/27/42-butterflies-of-north-america.html}

\supp{Dominikus Baur, Moritz Setfaner, and Rauref}{2013}{OECD Better Life index.}{OECD.}{https://www.oecdbetterlifeindex.org} \label{example:oecd}

\supp{Unknown}{2018}{Fewer Fireworks For Celebration, Better Air Quality To Start The New Year.}{ThePaper.cn}{https://image.thepaper.cn/html/zt/2019/02/fireworks/index.html}

\supp{Unknown}{2020}{Lifelines: A pandemic undertow.}{Periscopic Inc.}{https://lifelines.periscopic.com/}

\supp{Unknown}{2019}{Mapping police violence.}{}{https://mappingpoliceviolence.org/}

\supp{Pedro Cruz}{2013}{Um ecossistema político-empresarial.}{}{https://pmcruz.com/eco/}

\supp{Cathryn Ploehn, Kelsey Campbell, and Gayta Science}{2022}{Transgender day of remembrance.}{}{https://cathrynploehn.com/tdor/2022/}

\supp{Ivett Kovacs and Balázs Fekete}{2023}{HitStory Anthems.}{}{http://hitstoryproject.com/}

\supp{Nick Babich}{2024}{North America population.}{}{https://x.com/101babich/status/1742204064803230017}
\end{enumerate}

\subsubsection{Corpus II: Selected Static Metaphoric Visualizations}
\begin{raggedright}
\begin{sloppypar}
\begin{enumerate}[left=8pt]
    \renewcommand{\labelenumi}{[{\small M\arabic{enumi}}]}
    \setcounter{enumi}{20}

    \supp{Arshad Ejaz}{2021}{Entry-level jobs require 3+ years of experience.}{}{https://public.tableau.com/app/profile/arshad.ejaz/viz/Entry-leveljobsrequire3yearsofexperience_16304835725570/Dashboard}

\supp{Kimly Scott}{2023}{\#MomsWhoViz.}{}{https://public.tableau.com/app/profile/kimly.scott/viz/MomsWhoVizWomens HistoryMonth/MomsWhoVizWomensHistoryMonth}

\supp{Ivett Alexa}{2016}{A week of schedules}{}{https://public.tableau.com/app/profile/yvette/viz/Aweekofschedules/Aweekofschedules_1} \label{example:week}

\supp{Unknown}{2023}{Website user engagement flow.}{EVERSANA INTOUCH Data \& Analytics.}{https://www.informationisbeautifulawards.com/showcase/6176-website-user-engagement-flo}

\supp{Jeremy Wanner}{2023}{Rosa Pugilatu-The Glove Rose.}{INSEP.}{https://www.informationisbeautifulawards.com/showcase/6460-rosa-pugilatu-the-glove-rose}

\supp{Aoife O'Doherty}{2023}{Deep breaths in the city.}{}{https://www.behance.net/gallery/177988439/Deep-breaths-in-the-city}

\supp{Italo Doliva}{2023}{But what about Protein?}{}{https://www.informationisbeautifulawards.com/showcase/6325-but-what-about-protein}

\supp{Asha Daniels}{2023}{Amur leopards are disappearing.}{}{https://public.tableau.com/app/profile/asha4359/viz/AmurLeopardsareDisappearing/Dashboard1} \label{example:leopard}

\supp{Kimly Scott}{2023}{Liquid gold.}{}{https://public.tableau.com/app/profile/kimly.scott/viz/WorldBreastfeedingWeekLiquidGold/WorldBreastfeedingWeekLiquidGold}

\supp{Hardik Chandrahas}{2018}{Shifting gears: Visualizing cycle rides.}{National Institute Of Design.}{https://github.com/IllusionInk/personaldata}

\supp{Krisztina Szőcs}{2019}{Plot Parade.}{}{https://plotparade.com/}

\supp{Dea Bankova, Prasanta Kumar Dutta, Anurag Rao, and Aditi Bhandari}{2023}{A visual guide to Eurovision.}{Reuters.}{https://www.reuters.com/graphics/MUSIC-EUROVISION/dwpkdykkzvm/} \label{example:eurovision}

\supp{Unknown}{2017}{One angry bird: Emotional arcs of the past ten U.S. presidential inaugural addresses.}{Periscopic Inc.}{https://emotions.periscopic.com/inauguration/}

\supp{Alli Torban}{2018}{The woman of data viz.}{}{https://dataviztoday.com/shownotes/28}

\supp{Shirley Wu}{2016}{Dive fractals: Synchronized divings in the Olympics.}{}{https://shirleywu.studio/olympics/}

\supp{Stefanie Posavec}{2006}{Literary organism.}{}{https://www.stefanieposavec.com/shop-1}

\supp{Beautiful News Daily}{2018}{Oil spills have decreased.}{}{https://informationisbeautiful.net/beautifulnews/1276-oil-spills-decreased/}

\supp{Federica Fragapane}{2018}{One year of visual narratives.}{}{https://www.behance.net/gallery/69907239/One-Year-of-Visual-Narratives}

\supp{David McCandless, Miriam Quick, and Andrew Park}{2022}{Best in show: The ultimate data dog.}{}{https://informationisbeautiful.net/visualizations/best-in-show-whats-the-top-data-dog/}

\supp{Adam McCann}{2018}{History of Bruce Springsteen.}{}{https://duelingdatalarge.blogspot.com/2018/08/history-of-bruce-springsteen.html} \label{example:bruce}

\supp{Estadão}{2019}{Simulation shows which children are adopted (and which are not) in Brazil.}{}{https://arte.estadao.com.br/brasil/adocao/criancas/}

\supp{Abhilash}{2022}{What am I reading.}{}{https://github.com/projectpolymer/bookreading} \label{example:book}

\supp{Ali Tehrani}{2021}{Lack of racial diversity among winners at the Oscars.}{}{https://public.tableau.com/app/profile/ali15tehrani/viz/LackofRacialDiversityAmongWinnersattheOscars/Oscars}

\supp{Wendy Shijia}{2019}{McNutrition: The portraits of McDonald's food.}{}{https://public.tableau.com/app/profile/wendy.shijia/viz/McNutrition/McDonaldsNutrition} \label{example:clown}

\supp{Michela Lazzaroni}{2018}{The butterfly effect.}{}{https://www.behance.net/gallery/65433181/The-Butterfly-Effect-La-Lettura-336-dataviz} \label{example:butterfly}

\supp{Michela Lazzaroni}{2023}{The Poet's Journey: Visualizing Dante's Divine Comedy Characters.}{}{https://medium.com/make-your-data-speak/the-poets-journey-visualizing-dante-s-divine-comedy-characters-ed00433b7280}

\supp{Brendan Dawes}{2013}{Digital city portrait – London.}{}{https://brendandawes.com/projects/ee}

\supp{Sonja Kuijpers}{2017}{A view of despair.}{Studio Terp.}{https://www.studioterp.nl/a-view-on-despair-a-datavisualization-project-by-studio-terp/} \label{example:despair}

\supp{Neil Richards}{2020}{Premier league squads 2020/21.}{} {https://public.tableau.com/app/profile/neil.richards/viz/premierleagueflowers/premierleagueflowers?publish=yes} \label{example:squads}

\supp{Kie Ichikawa}{2021}{Memory Blossoms.}{}{https://medium.com/towards-data-science/data-viz-meets-death-a481498f7a58}

\end{enumerate}
\end{sloppypar}
\end{raggedright}

\subsection{Semi-structured Interview Template}
The following is the template for the semi-structured interview in the formative study.
\label{appendix:formative_template}
\begin{enumerate}[label=Q \negthinspace\arabic*., left=0pt]
    \item After viewing these examples (in Corpus I), can you think of the advantages and disadvantages of using animation in metaphoric visualization?
    \item Have you integrated animation design in your data visualization work? Could you elaborate on some examples?
    \begin{enumerate}[label*=\arabic*.]
        \item Based on [the introduced example], how do you implement animation?
        \item Have you encountered any difficulty in conceiving the [animation effects mentioned by the participant]?
        \item Do you find it challenging to come up with an animation design for your project?
    \end{enumerate}
    
    \item What is your motivation for adding animation in your past or ongoing projects? Do you see specific benefits?
    
    \item Under what situations will you opt not to use animation? Why?
    
    \item What is your general workflow when designing animation, such as [projects mentioned]? Does it come in parallel with the static appearance or follows a sequential order?

    \item Given your experience in creating general animation or animated visualizations, what is your expectation towards an animation tool for data visualizations? Do you have some pain points in mind to address? What are the most critical features you anticipate?
    
    \item Based on this design [selected example in Corpus II], can you identify areas of improvement through animation? Why is it good? Does this animation design significantly improve the quality of the work?
    \begin{enumerate}[label*=\arabic*.]
        \item Why/Why not add animations during [specific interaction afforded in the selected visualization]?
    \end{enumerate}
\end{enumerate}

\section{User Study}
\label{appendix:user}

\subsection{Self-developed Questionnaire}
\label{appendix:questionnaire}
Based on the statements below, users should provide self-ratings on a 7-point Likert scale (1: highly disagree, 4: neutral, 7: highly agree).
The items were mainly adapted from the System Usability Scale (SUS).
We adjusted the user selection in questions asked reversely to reflect the positiveness when reporting the results. The abbreviations marked in bold are not presented in the questionnaire.
\begin{itemize}
    \item \textbf{Simple abstraction:} I found the configuration of animation unnecessarily complex.
    \item \textbf{Easy to coordinate:} I found the coordination of visual elements unnecessarily complex.
    \item \textbf{Function integrity:} I found the various functions in this system were well integrated.
    \item \textbf{Consistency:} I thought there was too much inconsistency in DataSway.
    \item \textbf{Confidence in using:} I felt very confident using DataSway.
    \item \textbf{Learnability:} I would imagine that most people would learn to use this system very quickly.
    \item \textbf{Learning curve:} I needed to learn a lot of things before I could get going with this system.
    \item \textbf{Workflow Integration:} I found the system fits in my workflow.
\end{itemize}

\subsection{Post-Study Interview Template}
\label{appendix:post_interview}
The following is the template for the semi-structured interview in the user study.
The facilitator asks the participants these questions when they have finished the questionnaires.

\begin{enumerate}[label=Q \negthinspace\arabic*., left=0pt]
               \item Behavioral Inquiry
              \begin{itemize}
            \item You spent significant time on [observed activity]; why so?
        \end{itemize}
            \item Assumption Validation
                \begin{itemize}
                    \item Do you find the input format and output format reasonable?
                    \item Do you see other workflows that are not well-supported?
                \end{itemize}
            \item Perceived Pros
                \begin{itemize}
                    \item What part of DataSway do you like most? Why?
                    \item Can you compare the Chat-only interface and DataSway? Which do you prefer? Why?
                \end{itemize}
            \item Perceived Glitches
                \begin{itemize}
                    \item Do you find it challenging to communicate with LLMs? Which aspect?
                    \item What is the most unnatural setting in the system? How to amend it?
                    \item Do you have any particular complaints about the system?
                    \item Is there any conflict between your idea and what the system can offer?
                \end{itemize}
            \item Application Scenario
                \begin{itemize}
                    \item Combined with your previous experience, what stage in a project is the tool most beneficial?
                \end{itemize}
            \item Improvement Space
                \begin{itemize}
                    \item If you had a magic wand, what would you like the system to offer?
                \end{itemize}

\end{enumerate}